\documentclass[a4paper, 10pt]{article}
\usepackage[left=0.7 in, right=0.7in, top=1in, bottom = 1in]{geometry}

\usepackage{amsmath}
\usepackage{amssymb}
\usepackage{bbm}
\usepackage{graphicx}
\usepackage{amsthm}
\usepackage{color}
\usepackage{graphics}
\usepackage{epsfig}
\usepackage{cancel}
\usepackage{epstopdf}
\usepackage{algorithm}
\usepackage{algorithmic}
\usepackage{hyperref}
\usepackage{xcolor}
\usepackage{cleveref}
\usepackage{authblk}
\usepackage{tikz}
\usepackage{pgfplots}
\usepackage[normalem]{ulem}
\usepackage{soul,xcolor}
\usetikzlibrary{spy,calc}
\hypersetup{
colorlinks,
linkcolor={blue!100!black},
citecolor={blue!80!black},
urlcolor={blue!80!black}
}
\usepackage{natbib}

\newcommand{\mbf}[1]{\mathbf{#1}}

\newcommand{\mcal}[1]{\mathcal{#1}}
\newcommand{\bsym}[1]{\boldsymbol{#1}}

\newcommand{\wt}[1]{\widetilde{#1}}

\newcommand{\pd}[2]{\frac{\partial #1}{\partial #2}}

\DeclareMathOperator{\CGright}{\bsym{C}}
\DeclareMathOperator{\Dgrad}{\bsym{F}}

\DeclareMathOperator{\strPK}{\mbf{P}} 

\DeclareMathOperator{\Div}{\text{Div}}

\DeclareMathOperator{\p}{\partial}

\usepackage{array}
\newcolumntype{L}[1]{>{\raggedright\let\newline\\\arraybackslash\hspace{0pt}}m{#1}}
\newcolumntype{C}[1]{>{\centering\let\newline\\\arraybackslash\hspace{0pt}}m{#1}}
\newcolumntype{R}[1]{>{\raggedleft\let\newline\\\arraybackslash\hspace{0pt}}m{#1}}

\title{Instabilities in a compressible hyperelastic cylindrical channel due to internal pressure and external constraints}

\author[1]{{Sumit Mehta}}
\author[1]{{Gangadharan Raju}}
\author[2]{Shanmugam Kumar}
\author[2]{{Prashant Saxena}\thanks{Corresponding author email: prashant.saxena@glasgow.ac.uk}}

\affil[1]{Department of Mechanical and Aerospace Engineering \protect\\
Indian Institute of Technology Hyderabad, India}
\affil[2]{James Watt School of Engineering,  University of Glasgow, Glasgow G12 8LT, UK}

\date{}
\begin{document}
\maketitle
\numberwithin{equation}{section}

\begin{abstract} 
Pressurised cylindrical channels made of soft materials are ubiquitous in biological systems, soft robotics and metamaterial designs.
In this paper, we study large deformation and subsequent instability of a  thick-walled and compressible hyperelastic cylinder under internal pressure and external constraints.
The applied pressure can lead to elastic bifurcations along the axial or circumferential direction. 
Perturbation theory is used to derive the partial differential equations that govern the bifurcation behaviour of the cylindrical channel. Two cases of boundary conditions on the outer surface of the cylinder, namely, free and constrained are studied to understand their influence on the instability behaviour.
The derived equations are solved numerically using the compound matrix method to evaluate the critical pressure for instability. The effects of the thickness of the cylinder and the compressibility of the material on the critical pressure is investigated for both the boundary conditions.
The results reveal that for an isotropic material, the bifurcation occurs along the axial direction of the cylinder at lower critical pressure compared to circumferential direction for all cases considered herein. 
Finally, the tuneability of the bifurcation behaviour of transversely isotropic cylinder is demonstrated by considering reinforcements along the cylinder's axis, triggering bifurcation in the circumferential direction in certain cases. The findings of the study indicate that the instability-induced pattern formation would be useful for designing transforming material architectures such as soft robotics and soft metamaterials.
\end{abstract}

\section{Introduction}

\begin{figure}
\centering
 \includegraphics[width = 0.9\linewidth]{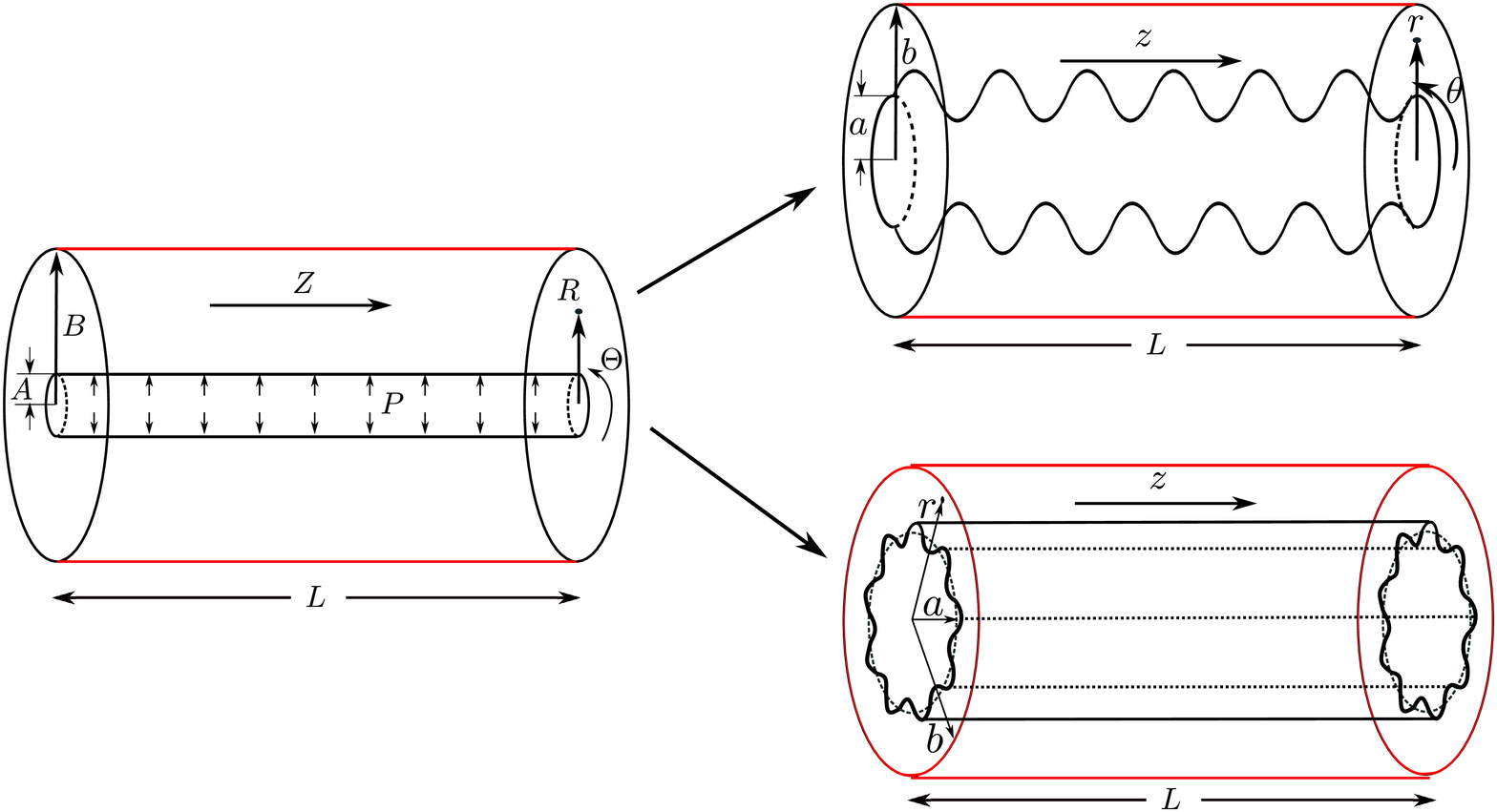}
 \caption{A long thick compressible cylindrical tube with internal radius $A$, external radius $B$ and length $L$ in the reference configuration that deforms to a cylinder with internal radius $a$ and external radius $b$ under an internal pressure and plane strain condition. The deformation can cause periodic patterns either (a) in the axial direction that maintains axisymmetry, or (b) in the circumferential direction that maintains the plane strain condition.} \label{fig: cylinder_instability}
 \end{figure}

Soft materials such as gels, soft tissues, and elastomers can undergo large deformation that can trigger elastic instabilities such as wrinkling and folding resulting in pattern formation \citep{barriere1996peristaltic, ciarletta2012peristaltic}.
The advantage of such materials is that they have high strength to modulus ratio and therefore can sustain large strain. 
Typically, they possess low elastic modulus which makes them prone to elastic instabilities such as wrinkling, creasing, and folding.
A cylindrical channel made  of soft hyperelastic material can undergo large deformation due to inflating pressure and can exhibit wrinkle patterns either along the circumferential or axial direction as shown in Figure \ref{fig: cylinder_instability}.
These undulating surface topographies are widely observed in biological systems such as skin, intestine, and mucus airways \citep{moulton2011circumferential}. 
Bifurcation of thin incompressible cylinder under inflating pressure is an extensively studied problem
\citep{haughton1979bifurcation1, benedict1979determination, fu2008post} in literature.
Thin-walled elastic tubes experience bulging and bending depending upon their length.
Bulging is dominant in short cylinders whereas long cylinders tend to bend when internal pressure is applied. 
On the contrary, a thick cylinder behaves in a different manner during  inflation. 
It first dilates homogeneously, then bifurcates and undulates either along the axial  \citep{cheewaruangroj2019peristaltic} or circumferential direction.
However, limited investigations have been undertaken to study bifurcation phenomenon in compressible solids experiencing large deformation \citep{Cai2019, Bakiler2021}.
Detailed analysis on stability and bifurcation of a compressible internally pressurised hyperelastic cylindrical structure is lacking and requires investigation.
Therefore, in this work, we study the circumferential and axial bifurcation phenomena in a pressurised compressible hyperelastic cylindrical channel as shown in the sketch in Figure \ref{fig: cylinder_instability}. 
We limit our discussion to only wrinkling instabilities by analysing linear perturbations to the principal deformation and have not considered creasing \citep{hong2009formation, Hohlfeld2011}, or folding \citep{tallinen2015mechanics, velankar2012swelling} phenomena which are also possible in soft solids.
             
Pressurised soft thick cylindrical channels are common surrogates to study biological systems such as blood flow through arteries \citep{hasan2015multilayered}, soft tissue deformation \citep{taghizadeh2015hyperelastic}, and have many clinical application such as biocompatible chips (organs on chips) and medical implants \citep{araci2014implantable, koh2016soft}.
Beyond these biomedical applications, soft channels also have important implication in 
metamaterials used for developing soft robotics \citep{rus2015design} such as soft grippers \citep{schumacher2015microstructures}. 
Soft microfluidic channels made of elastomer through soft lithography or rapid prototyping have been shown to be advantageous as deformation of these channels are useful in actuating the valves between the pumps \citep{unger2000monolithic}.
In addition, soft channels are encountered in polymeric hydrogels experiencing high strain and confined in granular medium for use as water reservoir in agriculture \citep{louf2021under}.

Bifurcation analysis of incompressible thick-walled tube under combined axial loading and external/internal pressure is discussed by \cite{haughton1979bifurcation2}. They studied the effect of wall-thickness which leads to non-homogeneous deformation. 
Recently, \cite{sang2016large} performed the stability analysis of incompressible rubber tube under internal pressure using Gent's strain energy function. \cite{anani2018stability} discussed the stability analysis of functionally graded incompressible thick-walled cylindrical and spherical shells using extended version of Ogden's strain energy function. The wall thickness has a significant influence on the stability of cylinder subjected to internal/external pressure. 
In particular, this motivates the investigation of the effect of displacement constraints along the outer surface, wall-thickness and material compressibility on critical pressure at which the instability occurs in the cylinder.

In the current work, we study large deformation of pressurised thick walled hyperelastic compressible cylinder and investigate the onset of instability under internal pressure. 
By incorporating the  compressible version of neo-Hookean constitutive model in the strain energy density function, the base state solutions are obtained for cylinders along azimuthal as well as axial direction. 
Both constrained and free boundary conditions are considered on the external surface of the cylinder.
The bifurcation solutions are then obtained by perturbing the principal solutions with a small parameter ($\epsilon$) using incremental deformation theory \citep{ogden1997non} along the circumferential and axial direction of the cylinder. 
The resulting incremental equations are solved numerically using the compound matrix method for computing critical value of inflating pressure.
The effect of cylinder thickness and material compressibility on the critical inflating pressure is also analysed.
The buckling modes corresponding to the critical pressure along the axial and circumferential directions are investigated. Finally, the influence of stiffening the cylindrical tube along the axial direction with fibre reinforcement and its role on the elastic instabilities is studied.

\subsection{Organisation of this manuscript}
The remainder of this paper is organised as follows.
In Section \ref{section_kinematics}, 
we discuss the base state solution for the cylinder subjected to internal pressure under free as well as constrained boundary conditions on the outer surface. 
In Section \ref{section_inc_theory}, we derive the incremental differential equations 
by perturbation in the circumferential and axial direction.
In Section \ref{results}, we derive the 
non-dimensional ordinary differential equations (ODEs) and evaluate the  
critical pressure that causes instability in circumferential as well as axial direction using compound matrix method and shooting method. 
Later in this section, we present a detailed discussion of numerical results also including the comparison of bifurcation solution in axial and circumferential direction.  
Finally, we conclude the work in Section \ref{section_conclusion} with the scope for potential future extensions.  
Supplementary mathematical derivations are given in the Appendix.
\subsection{Notation used in this manuscript}

\underline{Brackets}: Two types of brackets are used. Round brackets ( ) are used to define the functions applied on parameters or variables. Square brackets [ ] are used to clarify the order of operations in an algebraic expression.

\noindent \underline{Symbols}: A variable typeset in a normal weight font represents a scalar. 
A lower-case bold weight font denotes a
vector and bold weight upper-case font denotes a tensor.
Matrix of a tensor is depicted by enclosing the tensor in square brackets.
Tensor product of two second order tensors $\mbf{A}$ and $\mbf{B}$ is defined as either $[\mbf{A} \otimes \mbf{B}]_{ijkl} = [\mbf{A}]_{ij} [\mbf{B}]_{kl} $ or $[\mbf{A} \boxtimes \mbf{B}]_{ijkl} = [\mbf{A}]_{ik} [\mbf{B}]_{jl} $. 
Higher order tensors are written in bold calligraphic font with a superscript as $ \pmb{\mathcal{A}}^{(i)}$, where superscript `$i$' indicates that the function is differentiated $i+1$ times. For example, $\pmb{\mathcal{A}}^{(1)} =\displaystyle \frac{\partial^2 \Omega}{ \partial \mathbf{F} \partial \mathbf{F}} $ is a fourth order tensor. Operation of a fourth order tensor on a second order tensor is denoted as $ [\pmb{\mathcal{A}}^{(1)} : \mbf{A}]_{ij} = [\pmb{\mathcal{A}}^{(1)}]_{ijkl} [\mathbf{A}]_{kl}$.
Inner product is defined as $\mbf{A} \cdot \mbf{B} = [\mbf{A}]_{ij} [\mbf{B}]_{ij}$.
We use the word `Div' to denote divergence in three dimensions. The term $\delta \mbf{F} $ is used to represent the increment in $\mbf{F}$.

\noindent \underline{Functions}: $\det(\mathbf{F})$ denote the determinant of a tensor $\mathbf{F}$. $\text{tr}(\mathbf{F})$ denote the trace of a tensor $\mathbf{F}$.

\section{Kinematics and principal solution} \label{section_kinematics}
\begin{figure}[h]
 \centering
 \includegraphics[width=0.55\linewidth]{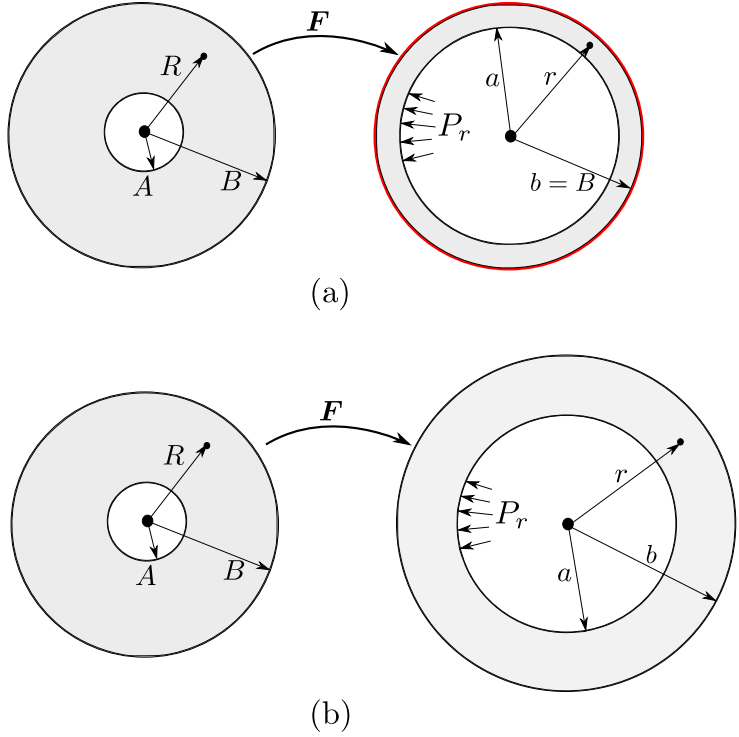}
 \caption{Cross-section of the cylinder in the reference and deformed configurations corresponding to the two boundary conditions considered.
 The inner and outer radii are $A$ and $B$ that transform to $a$ and $b$, respectively, due to an internal pressure $P_r$. (a) The outer surface is constrained forcing $b=B$. (b) The outer surface is free to expand.}
 \label{fig: kinematics}
\end{figure}

Consider an infinitely long thick cylinder with an internal radius $A$ and external radius $B$ in its stress-free reference configuration.
The cylinder is deformed by an internal pressure $P_r$ as shown in Figure \ref{fig: kinematics} under two types of boundary conditions (free and constrained) on the outer surface.
Let the cylindrical coordinates in the reference configuration be denoted by $(R, \Theta, Z)$ and in the deformed configuration by $(r, \theta, z)$.
In its deformed configuration, the internal radius of the cylinder is given by $a$ and the external radius is $b$. For the constrained boundary condition on the outer surface, $b=B$.
A plane strain problem is considered and therefore no dependence on the $Z$ coordinate is considered.
We also assume axisymmetry that removes any dependence on the $\Theta$ coordinate.
We denote the deformation gradient by $\Dgrad$ and the right Cauchy--Green deformation tensor as $\CGright = \Dgrad^{T} \Dgrad$.
For the current case of axisymmetric deformation, we can write the components of $\Dgrad$ in the cylindrical coordinate system as $ [\Dgrad] = \text{diag} (\lambda_r, \lambda_\theta, \lambda_z)$
where the principal stretch ratios can be written as
\begin{equation}
 \lambda_r = \pd{r}{R},  \quad \quad \lambda_\theta = \frac{r}{R}, \quad \quad \lambda_z = 1.
 \label{eq: lambda forms}
\end{equation}
The deformation function in the radial direction $r(R)$ is an unknown.

\subsection{Equilibrium and boundary conditions}
The balance of linear momentum 
\begin{equation}
 \Div \strPK = \mbf{0},
\end{equation}
can be written in cylindrical coordinates for this axisymmetric case with no dependence of variables along the $Z$ coordinate as
\begin{align}
 P_{Rr}' + \frac{1}{R} \left[ P_{Rr} - P_{\Theta \theta} \right] = 0. \label{eq: eqbm eqn 1}
\end{align}
Here, $\strPK$ is the first Piola--Kirchhoff stress tensor with components $P_{ij} := [\strPK]_{ij}$ and a prime denotes derivatives with respect to $R$.
There are no shear components of stress because of isotropy and $\mbf{F}$ being diagonal (axisymmetric deformation). 
For simplicity we use a compressible neo-Hookean energy density function for the hyperelastic material  \citep{limbert2018skin}
\begin{equation}
 \Omega (I_1, I_3) = \frac{\mu}{2} \big[ I_1 - 3 -  \, \text{log}\, I_3 \big] + \frac{\kappa}{4} \big[ \text{log}\, I_3 \big]^2 , \label{strain_energy}
\end{equation}
where the scalar invariants are defined as $I_1 = \text{tr}(\CGright), I_3 = J^2 = [\text{det} (\Dgrad)]^2$, $\mu$ is the ground state shear modulus, and $\kappa$ is a material parameter that relates to the ground state bulk modulus $K$ as $\kappa = K/2 - \mu/3$. 
Using \eqref{strain_energy}, the equilibrium equation \eqref{eq: eqbm eqn 1} is rewritten as (with detailed derivations in Appendix \ref{Inc_stress_app})

 \begin{align}
 \pd{}{R} \Bigg(  \alpha \bigg[ r' - \frac{1}{r'} \bigg] +  \frac{2}{r'} \text{log} \left( \frac{r r'}{R} \right) \Bigg) =
 \frac{\alpha}{R} \bigg[ \frac{r}{R} - r' \bigg] + \frac{1}{R} \left[ \alpha - 2   \text{log} \left( \frac{r r'}{R} \right) \right] \left[ \frac{1}{r'} - \frac{R}{r} \right]. \label{eq: 2nd ord ODE}
 \end{align}

This is a second order ODE for the unknown $r(R)$ with $R\in[A,B]$.
Note that here we have defined a dimensionless parameter $\alpha = \mu/ \kappa$.
In the linear elastic regime ($\Dgrad \approx \mbf{I}$), the parameter $\alpha$ is written in terms of the Poisson's ratio $\nu$ as $\alpha = (1-2\nu)/\nu$ 
which implies that for $\alpha = 0$, the cylinder is incompressible.
In order to assess the mechanical response for compressible cylinders, we perform computations for $\alpha > 0$.

\subsubsection{Constrained boundary conditions}
If the outer boundary of the cylinder is constrained as shown in Figure \ref{fig: kinematics}a, then the displacement boundary condition over the external surface is
 \begin{equation}
  r = B, \quad \quad \text{at} \quad \quad R=B ,
 \end{equation}
 and the traction boundary condition over the inner surface is
 \begin{equation}
  -P_r = P_{Rr}, \quad \quad \text{at} \quad \quad R = A,
 \end{equation}
 where $P_r$ is internal pressure. 
\subsubsection{Free boundary conditions}
If the outer boundary of the cylinder is free as shown in Figure \ref{fig: kinematics}b,
then the required traction boundary conditions are 
 \begin{equation}
  -P_r = P_{Rr}, \quad \text{at} \quad R = A, \quad  \text{and} \quad P_{Rr} = 0 \quad \text{at} \quad R=B.
 \end{equation}

 \subsection{Numerical solution for equilibrium}
 The second order ODE \eqref{eq: 2nd ord ODE} can be rewritten as a system of two first order ODEs by defining $ y_1 = r$ and $y_2 = r'$ as
 \begin{equation}
  \begin{bmatrix}
   1 & 0\\
   0 & \mcal{W}_1
  \end{bmatrix}
  \begin{bmatrix}
   y_1' \\
   y_2'
  \end{bmatrix}
  =
  \begin{bmatrix}
   y_2 \\
   \mcal{W}_2
  \end{bmatrix} , \label{eq: ODE system}
 \end{equation}
where the coefficients $\mcal{W}_1$ and $\mcal{W}_2$ in \eqref{eq: ODE system} are 
\begin{align}
\mcal{W}_1 & = \alpha \bigg[ 1 + \frac{1}{y_2^2} \bigg] + \frac{2}{y_2^2} \bigg[ 1 - \text{log} \left( \frac{y_1 y_2}{R} \right) \bigg] , \nonumber \\ 
\mcal{W}_2 & = \frac{\alpha}{R} \bigg[ \frac{y_1}{R} - y_2 \bigg] + \frac{1}{R} \Big[ \alpha - 2   \text{log} \left( \frac{y_1 y_2}{R} \right) \Big] \bigg[ \frac{1}{y_2} - \frac{R}{y_1} \bigg] + 2 \bigg[ \frac{1}{R y_2} - \frac{1}{y_1} \bigg].
\end{align}
 The corresponding boundary conditions transform to
 \begin{align}
 \alpha \bigg[ y_2 - \frac{1}{y_2} \bigg] +  \frac{2}{y_2} \text{log} \left( \frac{y_1 y_2}{R} \right) + \wt{P} = 0 , \quad \quad &\text{at} \quad R = A , \\
 y_1 = B, \quad \quad &\text{at} \quad R = B ,
 \end{align}
 for the constrained outer surface and
 \begin{align}
 \alpha \bigg[ y_2 - \frac{1}{y_2} \bigg] +  \frac{2}{y_2} \text{log} \left( \frac{y_1 y_2}{R} \right) + \wt{P} = 0 , \quad \quad &\text{at} \quad R = A , \\
 \alpha \bigg[ y_2 - \frac{1}{y_2} \bigg] +  \frac{2}{y_2} \text{log} \left( \frac{y_1 y_2}{R} \right) = 0, \quad \quad &\text{at} \quad R = B, \label{free_case_cond}
\end{align}
for the free outer surface of the cylinder. Here $\wt{P} = P_r/\kappa$ is the dimensionless internal pressure.

\begin{figure}
\centering
\includegraphics[width = 0.4\linewidth]{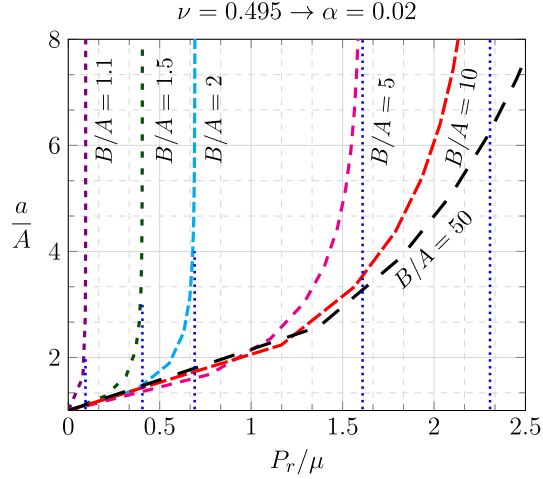}
\caption{The deformed dimensionless internal radius $a/A$ (dilation) as a function of the dimensionless applied internal pressure $P_r/\mu$ for a range of cylinder thickness ($B/A$) for a nearly incompressible cylinder $(\alpha = 0.02 \to \nu = 0.495)$. The plots are very close to those presented by \cite{cheewaruangroj2019peristaltic} for incompressible cylinders.} \label{Cheewaruangroj_fig_validation}
\end{figure}

In order to validate our current model, we compare the predictions with existing results for inflation of an incompressible cylinder with free boundary. 
Equations \eqref{eq: ODE system}--\eqref{free_case_cond} are solved using the \texttt{bvp4c} solver based on residual control in Matlab R2018a for $\alpha = 0.02$ (or $\nu = 0.495$) and for various cylinder thickness values, $B/A = 1.1,~ 1.5,...~ 50$. 
These results are presented in Figure \ref{Cheewaruangroj_fig_validation} and are in good agreement with the results reported by \cite{cheewaruangroj2019peristaltic} for incompressible cylinders ($\nu=0.5$).
The plots show the variation of the deformed inner radius $a/A$ (dilation) with respect to the normalised internal pressure. 
The contribution of $\kappa$ term in \eqref{strain_energy} is very small as $J \to 1$ or $\log(J) \to 0$ for the parameter, $\alpha = 0.02$.
We note the existence of a critical pressure, $P_r = \mu \log(B/A)$ at which the divergence happens leading to cavitation phenomenon (blue dotted line in Figure \ref{Cheewaruangroj_fig_validation}). This is not observed as $B/A$ tends to infinity.

\begin{figure}
\centering
\begin{tabular}{c c c}
\includegraphics[width=0.3\linewidth]{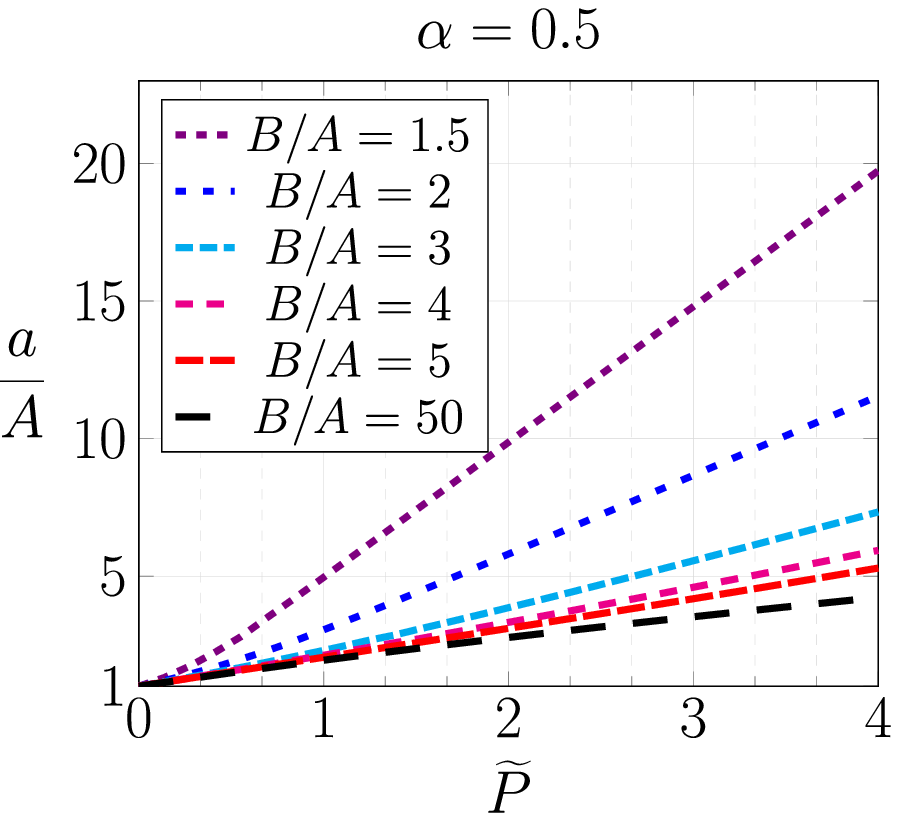}
&
\includegraphics[width=0.3\linewidth]{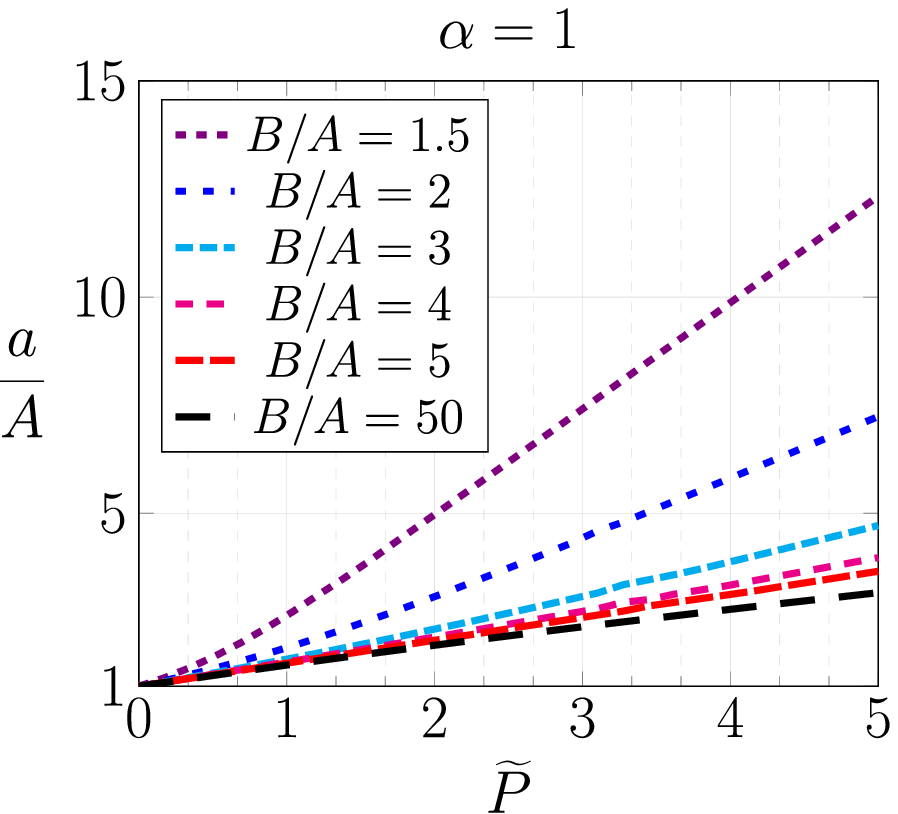}
&
\includegraphics[width=0.3\linewidth]{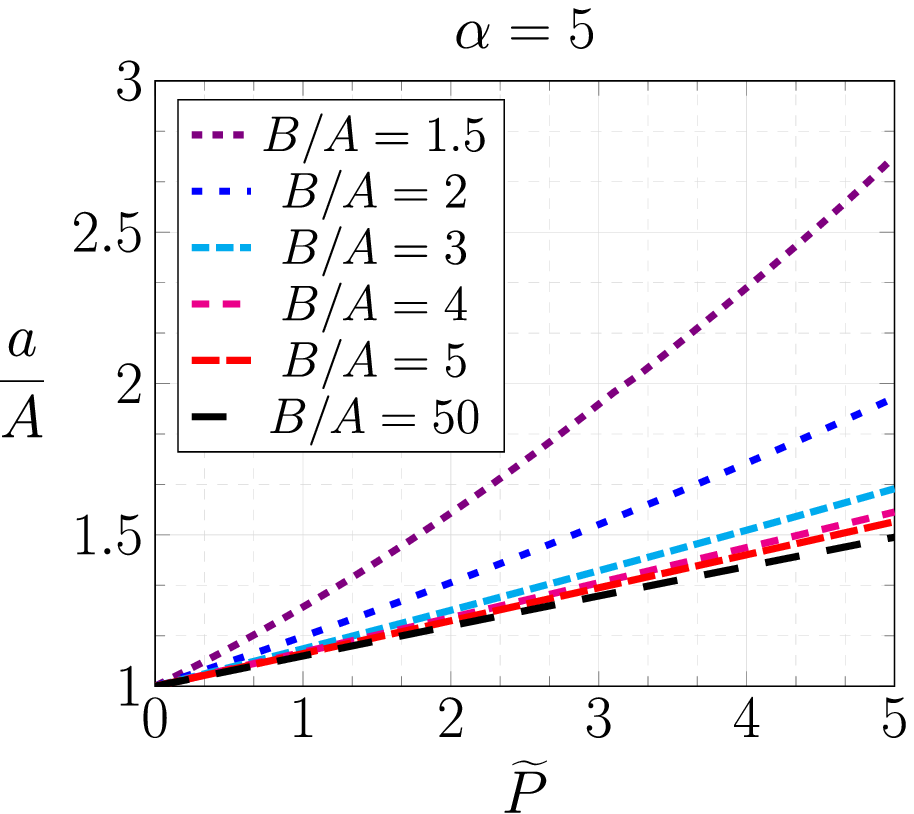}
\\
(a) & (b) & (c)
\end{tabular}
\caption{Free inflation: Variation of the deformed internal radius $a/A$ with the applied internal pressure $\widetilde{P}$ for different values of radius ratio ($B/A$) and material parameter (a)~$\alpha = 0.5$ (b)~$\alpha=1$, (c)~$\alpha =  5$.} \label{Free_pre_buckling}
\end{figure}

\begin{figure}
\centering
\begin{tabular}{c c c}
\includegraphics[width=0.29\linewidth]{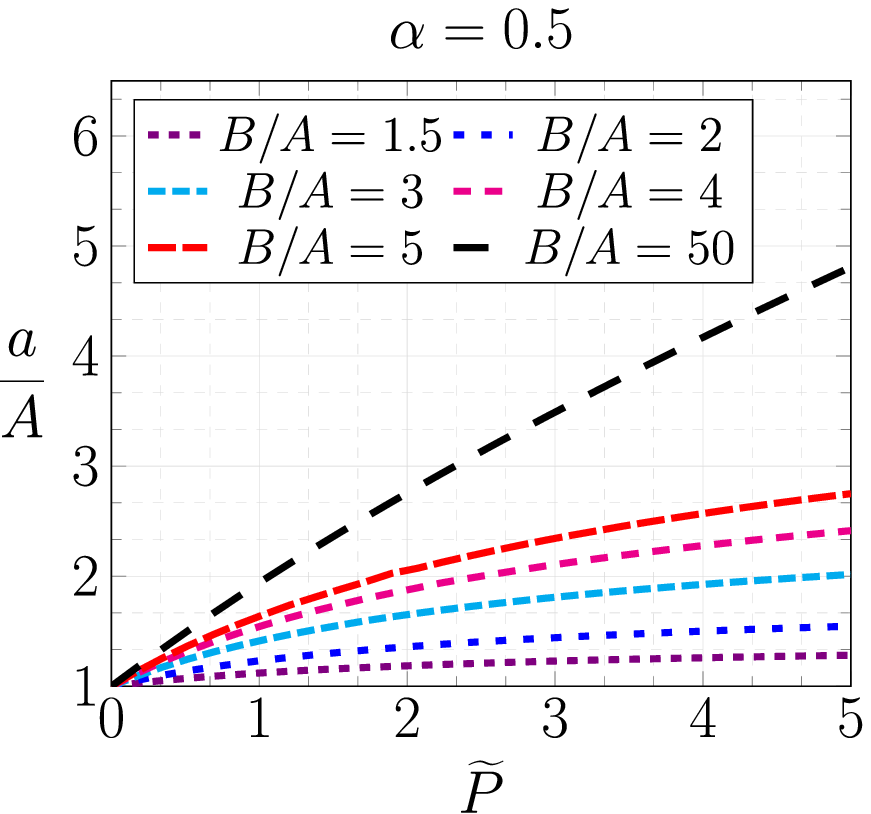}
&
\includegraphics[width=0.3\linewidth]{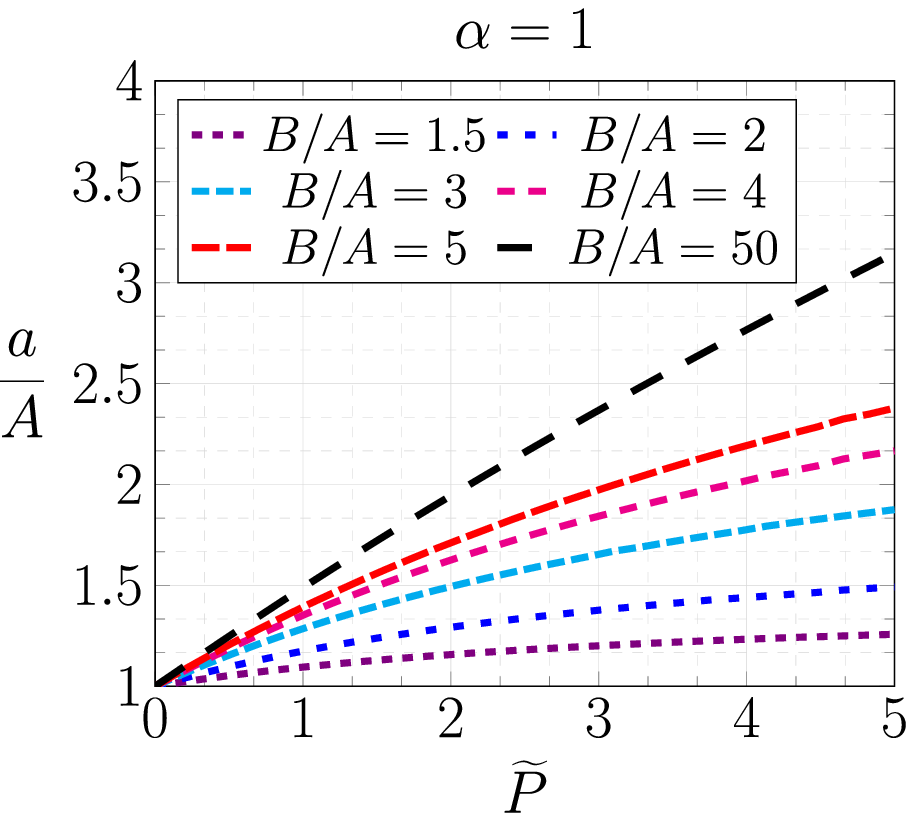}
&
\includegraphics[width=0.3\linewidth]{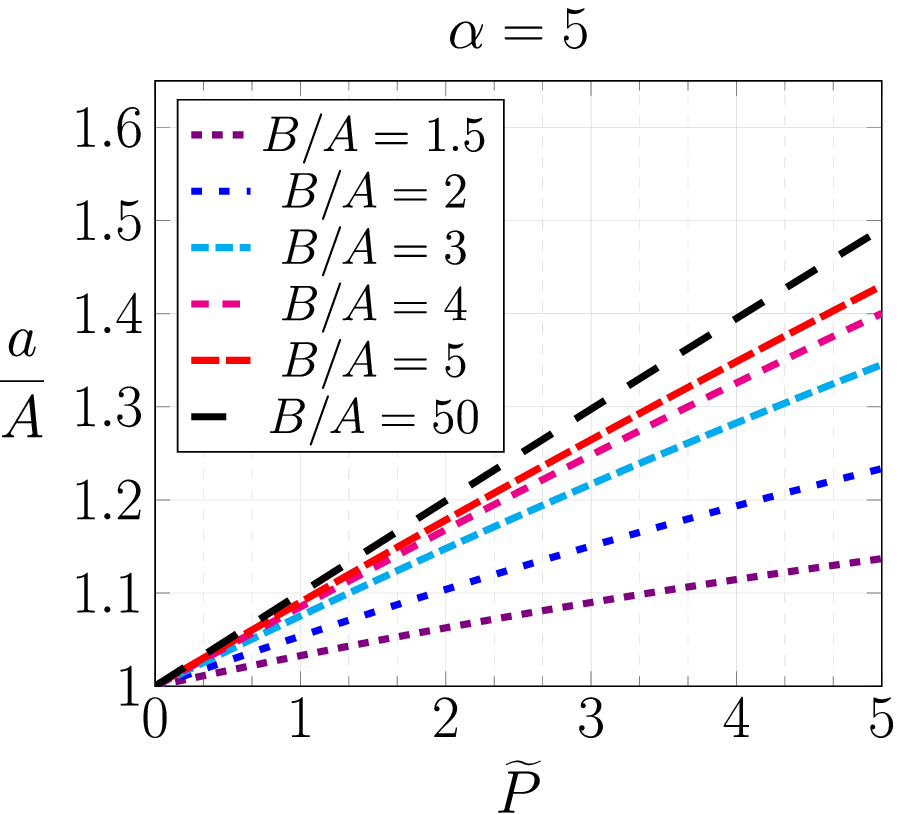}
\\
(a) & (b) & (c)
\end{tabular}
\caption{Constrained inflation: Variation of the deformed internal radius $a/A$ with the applied internal pressure $\widetilde{P}$ for different values of radius ratio ($B/A$) and material parameter (a)~$\alpha = 0.5$ (b)~$\alpha=1$, (c)~$\alpha = 5$.}  \label{const_pre_buckling}
\end{figure}

Results for the deformation of compressible cylinders with free and constrained external boundaries are shown in Figure \ref{Free_pre_buckling} and Figure  \ref{const_pre_buckling}, respectively.
The plots show the variation of the deformed inner radius $a/A$ with the internal applied pressure $\widetilde{P}$ for different values of the material parameter $\alpha$.

Results of constrained and unconstrained cases show that the maximum dilation for any value of $B/A$ decreases with increasing value of $\alpha$. 
In the constrained case, the inner radius deforms less for smaller value of radius ratio $(B/A)$ and this is due to the fixed boundary which causes resistance to dilation. 
As the wall thickness increases, the deformation of inner radius increases due to less dilation resistance from the boundary constraints. 
For the constrained case, the variation of $a/A$ with respect to internal pressure is nonlinear for less value of $\alpha$ and becomes linear for high values of $\alpha$.
This trend is markedly opposite in the unconstrained case shown in Figure \ref{Free_pre_buckling} due to the stress free boundary and the variation of $a/A$ is almost linear for all values of $\alpha$. This behaviour is in contrast to the nonlinear variation observed for the nearly incompressible case in Figure \ref{Cheewaruangroj_fig_validation}.
Here, the thick cylinder deforms less as compared to thin cylinder for same amount of pressure and material/geometrical parameters. 

When the limit $B/A \to \infty$, it corresponds to a cylindrical channel in an infinite space for which the influence of boundary is negligible and the deformation in the cylinder is identical for both constrained and unconstrained cases. We demonstrate this by choosing $B/A =50 $ in the simulations and it is observed that these results corresponding to lower bound for unconstrained cases and upper bound for constrained case converge in Figures \ref{Free_pre_buckling} and \ref{const_pre_buckling}.

\section{Incremental equations} \label{section_inc_theory}
In this section, we derive the partial differential equations that govern the instability behaviour of cylindrical channels subjected to internal pressure based on incremental theory.
We apply small perturbations to the primary deformation $(r, \theta, z)$ scaled by a parameter $0< \epsilon \ll 1$ such that the total deformation is
\begin{equation}
    \hat{r} = r + \epsilon u, \quad \hat \theta = \theta + \epsilon v, \quad \text{and} \quad \hat{z} = z + \epsilon w,
\end{equation}
and the associated deformation gradient tensor is 
\begin{align}
\mathbf{F}+ \delta \mbf{F}=\begin{bmatrix}
\displaystyle \frac{\partial \hat{r}}{\partial R} & \displaystyle \frac{1}{R} \frac{\partial \hat{r}}{\partial \Theta} & \displaystyle \frac{\p \hat{r}} {\p Z} \\[7pt]
\hat{r} \displaystyle \frac{\partial \hat{\theta}}{\partial R} & \displaystyle \frac{\hat{r}}{R}\frac{\partial \hat{\theta}}{\partial \Theta} & \hat{r} \displaystyle \frac{\p \hat{\theta}}{\p Z} \\[7pt]
 \displaystyle \frac{\p \hat{z}}{\partial R} & \displaystyle \frac{1}{R}\frac{\partial \hat{z}}{\partial \Theta} & \displaystyle \frac{\p \hat{z}}{\p Z}
\end{bmatrix}.
\label{deformation_grad_new}
\end{align}
Here, $\delta \mbf{F}$ is the incremental deformation gradient tensor.
The incremental first Piola--Kirchhoff stress tensor \citep{ogden1997non} is then obtained as
\begin{align}
\delta \mathbf{P} = \pmb{\mathcal{A}}^{(1)}\delta \mathbf{F} + \frac{1}{2}\pmb{\mathcal{A}}^{(2)}[\delta \mathbf{F}, \delta \mathbf{F}]+... \ , \label{Incremental_stress}
\end{align}
where $\pmb{\mathcal{A}}^{(i)}= \displaystyle\frac{\partial^{i+1} \Omega}{\partial \mathbf{F}^{i+1}}$ are the elastic moduli of the material. The first order modulus is 
\begin{equation}
\begin{aligned}
\pmb{\mathcal{A}}^{(1)}  & = \frac{\partial^2 \Omega}{\partial \mathbf{F} \partial \mathbf{F}} = \mu \bigg[\mathbb{I} - \mathbb{T}[-\mathbf{F}^{-1} \boxtimes \mathbf{F}^{-T}]\bigg] + 2 \kappa \bigg[ \mathbf{F}^{-T} \otimes \mathbf{F}^{-T} + [\log J] \mathbb{T}[-\mathbf{F}^{-1} \boxtimes \mathbf{F}^{-T}]\bigg],
\end{aligned} \label{elasti_modulii}
\end{equation}
where $[\mathbb{I}]_{ijkl} = \delta_{ij} \delta_{kl}$ and $[\mathbb{T}]_{ijkl} = \delta_{il} \delta_{jk}$.
Upon ignoring the higher order terms in \eqref{Incremental_stress}, the incremental first Piola--Kirchhoff stress tensor is given as 
\begin{align}
\delta \mathbf{P} = \mu\bigg[ \delta \mathbf{F} + {[\mathbf{F}^{-1} [\delta \mathbf{F}]~ \mathbf{F}^{-1}]^{T}} \bigg] + 2 \kappa \bigg[ \mathbf{F}^{-T}~ \text{tr}(\mbf{F}^{-1} [\delta \mathbf{F}]) - \log J {[\mathbf{F}^{-1} [\delta \mathbf{F}]~ \mathbf{F}^{-1}]^{T}} \bigg]. \label{inc_stress}
\end{align}
Balance of traction in the current configuration subjected to internal pressure is 
\begin{align}
\pmb{\sigma} \mbf{n} = -P_r \mbf{n},
\end{align}
where $\pmb{\sigma}$ is the Cauchy stress tensor, $P_r$ is the internal pressure and $\mathbf{n}$ is the unit outward normal in the current configuration. 
This can be rewritten in the reference configuration as
\begin{equation}
\begin{aligned}
\mbf{PN} = -J P_r \mbf{F}^{-T} \mbf{N},
\end{aligned} \label{trans_cauchy_to_Piola}
\end{equation}
where $\mathbf{N}$ is the unit outward normal in the reference configuration. Using transformation \eqref{trans_cauchy_to_Piola}, the incremental equilibrium equation and the associated incremental boundary conditions are 
\begin{subequations} \label{equi_inc_stress}
\begin{align}
\text{Div} (\delta \mbf{P}) &= \mbf{0}, \label{inc_eqbm_eq}\\
[\delta \mbf{P}] \mbf{N} &=  J P_r \mbf{F}^{-T} [\delta \mbf{F}]^{T} \mbf{F}^{-T} \mbf{N}  - J P_r \text{tr} \Big(\mbf{F}^{-1} [\delta \mbf{F}] \Big) \mbf{F}^{-T} \mbf{N}. \label{bc_for_inc_stress}
\end{align}
\end{subequations}
The detailed mathematical derivations associated with equations \eqref{elasti_modulii} -- \eqref{equi_inc_stress} are presented in Appendix \ref{Inc_stress_app}.
In this work, we seek two types of bifurcation from the primary solution.
The first one is a solution that satisfies the plane strain condition $(w=0)$ and causes perturbations in the radial-circumferential direction (i.e., $r,\theta$ coordinates).
The second bifurcation problem is the perturbation of the solution along the radial-axial direction (i.e., $r,z$ coordinates) and no variation along the circumferential coordinate, that is, $v=0$. The bifurcation along the axial direction is also possible by perturbing the primary solution only along radial component of the cylinder i.e., $v=w=0$ in contrast to the latter case of bifurcation.

\subsection{Perturbation along the circumferential direction} \label{subsec_circum_eqn}
We first {apply small perturbations to the principal solution by choosing} $0< \epsilon \ll 1$  which satisfy the plane strain condition such that,
\begin{align}
\hat{r} (R,\Theta) = r(R) + \epsilon u(R, \Theta), \hspace{0.5in} \hat{\theta}(R,\Theta) =\Theta + \epsilon v(R,\Theta),
\end{align}
where $r = r(R)$, $\theta = \Theta$ are the primary solution and ($\hat{r}, \hat{\theta}$) represent the deformation function upon perturbation.
The associated two-dimensional deformation gradient and its increment are  
\begin{align}
\mathbf{F} = 
\begin{bmatrix}
\lambda_r & 0 \\
0 & \lambda_\theta
\end{bmatrix}, \qquad \delta \mbf{F} = \begin{bmatrix}
\displaystyle \frac{\partial u}{\partial R} & \displaystyle \frac{1}{R} \frac{\partial u}{\partial \Theta} \\[10pt]
\displaystyle r \frac{\partial v}{\partial R} & \displaystyle \frac{r}{R} \frac{\partial v}{\partial \Theta}
\end{bmatrix}. \label{def_grad}
\end{align} 
Consider a sinusoidal perturbation as an ansatz
\begin{equation}
\begin{aligned}
u(R,\Theta)=  \Delta f(R) \cos(n \Theta), \qquad \text{and} \qquad 
v(R,\Theta)= \Delta g(R) \sin(n \Theta), 
\end{aligned} \label{ansatz1}
\end{equation}
where `$n$' denotes the wave number in circumferential direction. On substituting \eqref{ansatz1} in the equilibrium equation \eqref{inc_eqbm_eq} and collecting only $O(\epsilon)$ terms, we obtain the incremental differential equations for the functions $\Delta f$ and $\Delta g$ as
\begin{align}
&-r' r^2 R^2 \big[-\alpha r'^2 + 2 \log \left( \frac{r r'}{R}\right) - 2 - \alpha \bigg] \Delta f'' \nonumber\\
 & \hspace{0.5in}+ r R \bigg[ \big[2 r'' r R - r'r \big] 2  \log \left( \frac{r r'}{R}\right) + r'^3 r \alpha - 6 r'' r R  - 2 r'' r R \alpha - 2 r'^2 R  + 4 r' r  + r' r \alpha \bigg] \Delta f'\nonumber \\
&  \hspace{0.5in} + r' \bigg[ -r'^2 r^2 \alpha n^2 + r'^2 R^2 2  \log \left( \frac{r r'}{R}\right) - 2 r'' r R^2  - r'^2 r^2 \alpha - 4 r'^2 R^2  - r'^2 R^2 \alpha + 2 r' r R  \bigg] \Delta f \nonumber \\
& \hspace{0.5in}  -r'^2 r^2 R^2 n \bigg[2  \log \left( \frac{r r'}{R}\right) - 2  - \alpha \bigg] \Delta  g' \nonumber\\
& \hspace{0.3in}+ r' r n \bigg[ 2  \log \left( \frac{r r'}{R}\right) r'^2 R^2 - 2 r'' r R^2  - r'^2 r^2 \alpha - 2 r'^2 R^2 - r'^2 R^2 \alpha + 2 r'r R \bigg] \Delta  g =0, \label{diff_after_comp_eps_1}
\end{align}
\begin{align}
& \bigg[r'^2 r^2 R^2 \alpha \bigg] \Delta  g'' 
 -r' r R \bigg[ -r'^2 R \alpha + 2  \log \left( \frac{r r'}{R}\right) R - r' r \alpha - R \alpha \bigg] \Delta  g' \nonumber\\
& r'^2 n^2 \bigg[ 2  \log \left( \frac{r r'}{R}\right) R^2 - r^2 \alpha - 2 R^2  - R^2 \alpha \bigg] \Delta  g +
 r' R^2 n \bigg[ 2  \log \left( \frac{r r'}{R}\right) - 2  -\alpha \bigg] \Delta f'\nonumber \\
& -n \bigg[ \big[r'' R^2 - r' R \big] 2 \log \left( \frac{r r'}{R}\right) - 2 r'' R^2  - r'' R^2 \alpha + 2 r'^2 r \alpha + 2 r' R  + r' R \alpha \bigg] \Delta f =0, \label{diff_after_comp_eps_2}
 \end{align}
and the associated boundary condition \eqref{bc_for_inc_stress} is rewritten as
\begin{align}
\begin{bmatrix}
\delta P_{rR} & \delta P_{r \Theta}  \\ 
\delta P_{\theta R} & \delta P_{\theta \Theta} \end{bmatrix}
\begin{bmatrix}
1\\
0
\end{bmatrix}
 = 
J P_r \mbf{F}^{-T} [\delta \mbf{F}]^{T} \mbf{F}^{-T} 
\begin{bmatrix}
1\\
0
\end{bmatrix}
 - J P_r \text{tr}(\mbf{F}^{-1} \delta \mbf{F}) \mbf{F}^{-T} \begin{bmatrix}
1\\
0
\end{bmatrix}.  \label{traction_bc}
\end{align}
The inner and outer boundary conditions for constrained cylinder are derived by collecting the linear order terms in $\epsilon$
\begin{subequations} \label{bc_1}
\begin{equation} 
\left.
\begin{aligned}
\big[2 r' A \kappa\big] \Delta f - r \bigg[ -r'^2 A \mu +  A \left[ 2 \kappa \log \left( \frac{r r'}{A}\right)  \right] +  r r' P_r - 2 A \kappa -  A  \mu \bigg] \Delta f' \\
+  [2 r r' A \kappa n] \Delta g = 0, \\
n  \bigg[A \left[ 2 \kappa \log \left( \frac{r r'}{A} \right) \right] + r r' P_r - A \mu \bigg] \Delta f  + [A r^2 r' \mu] \Delta g = 0,
\end{aligned} \right\} \quad \text{at} ~R = A,
\end{equation}
\begin{equation}
\Delta f = \Delta g = 0 \qquad \qquad \qquad  \text{at} ~R = B,
\end{equation} 
\end{subequations}

\subsection{Perturbation along the axial direction} \label{subsec_axial_eqn}

In this section, we {apply small increments ($ 0 < \epsilon \ll 1$) to} the principal solution with perturbations along the axial direction satisfying axisymmetry such that
\begin{align}
\hat{r} = r(R) + \epsilon ~ U(R, Z), \qquad \hat{\theta}= \Theta,  \qquad \hat{z} = Z + \epsilon ~ W(R,Z). \label{increment_axial_dir}
\end{align} 
The deformation gradient 
and the corresponding incremental deformation gradient tensor are obtained  by collecting $O(\epsilon)$ terms as 

\begin{align}
\mathbf{F}
=
\begin{bmatrix}
\lambda_r & 0 & 0\\
0 & \lambda_\theta & 0\\
0  & 0 & 1
\end{bmatrix}, \qquad \delta \mbf{F} = \begin{bmatrix}
\displaystyle \frac{\partial U}{\partial R} & 0 & \displaystyle  \frac{\partial U}{\partial Z} \\[7pt]
0 & \displaystyle  \frac{U}{R} & 0\\[7pt]
\displaystyle \frac{\partial W}{\partial R} & 0 & \displaystyle \frac{\partial W}{\partial Z}
\end{bmatrix},
\end{align}
where, $r = r(R),~ \theta = \Theta, ~ z=Z $ is the primary solution. 
We consider the following ansatz
\begin{equation} \label{ansatz2}
\begin{aligned}
U(R,Z)&=  \Delta \bar{f}(R) \cos \left(m \frac{2 \pi}{L} Z \right), \quad \text{and} \quad W(R,Z)= \Delta \bar{h}(R) \sin \left(m \frac{2 \pi}{L} Z \right). 
\end{aligned}
\end{equation}
Here, `$m$' represents the wavenumber along the axial direction. We take the analysis domain in Z direction as $0<Z<L$, where $L$ is the length of the cylinder. Upon substituting \eqref{ansatz2} in the equilibrium equation \eqref{inc_eqbm_eq} and collecting $O(\epsilon)$ terms, we obtain the incremental ODEs for $\Delta \bar{f}$ and $\Delta \bar{h}$ as
\begin{equation}\label{first_order_diff_eq_axial}
\begin{aligned}
\Delta {\bar{f}}''&=-\frac{1}{c_1}\bigg[c_2 \Delta {\bar{f}}' + c_3 \Delta \bar{f} +c_4 \Delta {\bar{h}}' + c_5 \Delta \bar{h} \bigg], \\
\Delta {\bar{h}}''&=-\frac{1}{d_1} \bigg[d_2 \Delta {\bar{h}}' + d_3 \Delta \bar{h} + d_4 \Delta {\bar{f}}' + d_5 \Delta {\bar{f}} \bigg],
\end{aligned}
\end{equation} 
where 
\begin{align*}
c_1 &= r' r^2 R^2 \bigg[ r'^2 \alpha - 2 \log \left(\frac{r r'}{R}\right) + 2 + \alpha \bigg],\\
c_2 &= r R \bigg[ r'^3 r \alpha + [2 r r'' R - r' r]~ 2 \log \left(\frac{r r'}{R}\right) -  2 r'^2 R - 6 r r'' R - 2 r r'' R \alpha + 4 r' r + r' r  \alpha  \bigg], \\
c_3 & = -r' \bigg[ r'^2 r^2 R^2 \alpha \left[m \frac{2 \pi}{L}\right]^2 - 2 \log \left(\frac{r r'}{R}\right) r'^2 R^2  + r'^2 r^2 \alpha + 4 r'^2 R^2 + r'^2 R^2 \alpha + 2 r r'' R^2 - 2 r' r R \bigg],\\
c_4 & = -r'^2 r^2 R^2 m \frac{2 \pi}{L} \bigg[2 \log \left( \frac{r r'}{R} \right) - 2 -\alpha \bigg], \qquad
c_5 = -2 r' r R m \frac{2 \pi}{L} \bigg[ r'^2 R + r r'' R - r' r \bigg].
\end{align*}
\begin{align*}
d_1 &= r'^2 R \alpha, \qquad
d_2  = r'^2 \alpha, \qquad
d_3  = \left[m \frac{2 \pi}{L}\right]^2 R r'^2 \bigg[ 2 \log \left(\frac{r r'}{R}\right) - 2 - 2 \alpha \bigg], \\
d_4 & = r' R m \frac{2 \pi}{L} \bigg[ 2 \log \left(\frac{r r'}{R}\right) - 2 - \alpha \bigg], \nonumber \\
d_5  &=   m \frac{2 \pi}{L} \bigg[ [r' - r'' R] ~ 2 \log \left(\frac{r r'}{R}\right)  + 2 r'' R + r'' R \alpha - 2 r' - r' \alpha \bigg].
\end{align*} 
The boundary condition \eqref{bc_for_inc_stress} for constrained cylinder is given by 
\begin{subequations} \label{axial_constrained_bc}
\begin{equation}
\left.
\begin{aligned}
c_{11} \Delta \bar{f} + c_{22} \Delta {\bar{f}}' + c_{33} \Delta \bar{h} &= 0, \\
d_{11} \Delta \bar{f} + d_{44} \Delta \bar{h}' &= 0,
\end{aligned} \right\} \qquad \text{at} \quad R=A,
\end{equation}
\begin{align}
\Delta \bar{f} = \Delta \bar{h} &= 0 \qquad  \text{at} \qquad R=B,
\end{align}
\end{subequations}
where the coefficients are defined as
\begin{align*}
c_{11} &=  2 r' A, \qquad
c_{22}  =  -r \bigg[ -r'^2 A \alpha + 2 \log \left(\frac{r r'}{A}\right) A + r \wt{P} r' - 2 A - A \alpha \bigg], \\
c_{33} & = 2 r r' A m \frac{2 \pi}{L}, \qquad
d_{11}  = m \frac{2 \pi}{L} \bigg[ 2 \log \left(\frac{r r'}{A}\right) A + r \wt{P} r' - A \alpha \bigg], \qquad
d_{44}  = \alpha r' A.
\end{align*}
\subsubsection{Perturbation only along radial component}
In this case, we apply small increments ($0< \epsilon <<1$) to the principal solution considering $W =0$ in contrast to \eqref{increment_axial_dir}. We seek the bifurcation solution in the axial direction of a cylinder by perturbing only radial component using the following ansatz 
\begin{equation} \label{ansatz3_one_dim}
\begin{aligned}
\widetilde{r}(R,Z)&= r(R) + \epsilon \Delta \widetilde{f}(R) \cos \left(\widetilde{m} \frac{2 \pi}{L} Z \right), \quad \widetilde{\theta} = \Theta, \quad \text{and} \quad \widetilde{z} = Z, 
\end{aligned}
\end{equation}
where ($\widetilde{r}, \widetilde{\theta}, \widetilde{z})$ denotes the incremental cylindrical coordinates in the deformed configuration. Eq. \eqref{ansatz3_one_dim} is attributed to the presence of only radial strain which resists the applied pressure in the axial bifurcation case. However, in Section \ref{subsec_circum_eqn}, the applied internal pressure is resisted by radial as well as the circumferential strain in the cylinder.
Here, on substituting \eqref{ansatz3_one_dim} in \eqref{inc_eqbm_eq} and collecting the first order $\epsilon$ terms, we obtain the incremental ODE for the function $\Delta \widetilde{f}$ alone as
\begin{align}
&r' r^2 R^2 \bigg[ r'^2 \alpha - 2 \log \left(\frac{r r'}{R}\right) + 2 + \alpha \bigg] \Delta \bar{f} \nonumber\\
& + r R \bigg[ r'^3 r \alpha + [2 r r'' R - r' r]~ 2 \log \left(\frac{r r'}{R}\right) -  2 r'^2 R - 6 r r'' R - 2 r r'' R \alpha + 4 r' r + r' r  \alpha  \bigg] \Delta \widetilde{f} \nonumber \\
& - r' \bigg[ r'^2 r^2 R^2 \alpha \left[\widetilde{m} \frac{2 \pi}{L}\right]^2 - 2 \log \left(\frac{r r'}{R} \right) r'^2 R^2 + r'^2 r^2 \alpha + 4 r'^2 R^2 + r'^2 R^2 \alpha + 2 r r'' R^2 - 2 r' r R \bigg] \Delta \widetilde{f} = 0.
\end{align}
The inner and the outer surface boundary conditions \eqref{bc_for_inc_stress} for the constrained cylinder are
\begin{subequations}
\begin{align}
&[2 r' A] \Delta \widetilde{f}' - r \bigg[ -r'^2 A \alpha + 2 \log \left(\frac{r r'}{A}\right) A + r \wt{P} r' - 2 A - A \alpha \bigg] \Delta f = 0, \quad \text{at} \quad R = A,\\
& \Delta \widetilde{f} = 0, \quad \text{at} \quad R = B,
\end{align}
\end{subequations}  
and the boundary conditions for the  unconstrained cylinder are
  \begin{subequations}
\begin{align}
&[2 r' A] \Delta \widetilde{f}' - r \bigg[ -r'^2 A \alpha + 2 \log \left(\frac{r r'}{A}\right) A + r \wt{P} r' - 2 A - A \alpha \bigg] \Delta \widetilde{f} = 0, \quad \text{at} \quad R = A,\\
&[2 r' B] \Delta \widetilde{f}' - r \bigg[ -r'^2 B \alpha + 2 \log \left(\frac{r r'}{B}\right) B - 2 B - B \alpha \bigg]  \Delta \widetilde{f} = 0, \quad \text{at} \quad R = B.
\end{align}
\end{subequations} 

\section{Numerical solution and discussion} \label{results}
The ODEs derived in Sections \ref{subsec_circum_eqn} -- \ref{subsec_axial_eqn} are reformulated in Appendix \ref{reform_eqn_app} for ease of numerical solution.
We compute the numerical solution using a shooting method \citep{haughton1979bifurcation2, saxena2018finite} as well as the compound matrix method \citep{haughton1997eversion, haughton2008evaluation, Mehta2021}.
A detailed explanation of the compound matrix method and shooting method with associated mathematical equations is given in Appendix \ref{numerical_tech_app}.

\subsection{Comparison of the numerical schemes}
Shooting method and the compound matrix method are implemented in the Matlab 2018a programming environment. The \texttt{ode45} ODE solver that implements an explicit Runge--Kutta method and \texttt{fminsearchbnd} optimisation subroutine \citep{JohnDErrico2021} 
based on Nelder--Mead simplex algorithm is used.
A tolerance value of $10^{-8}$ is chosen to compute the bifurcation solution.
Both the methods compute the same results, but the
compound matrix method is approximately three times faster than the shooting method.
As an example, on a computer with an $8$ core, $2.10$ GHz processor and $48$ GB of RAM, computation of the curve corresponding to $\alpha=1, n=1$ in Figure \ref{var_wall_press_alpha_1_and_5_circum}a takes $90$ seconds using the compound matrix method and $337$ seconds using the shooting method.  

\subsection{Bifurcation of solution for a constrained cylinder}

\begin{figure}
\centering
\includegraphics[width=0.9\linewidth]{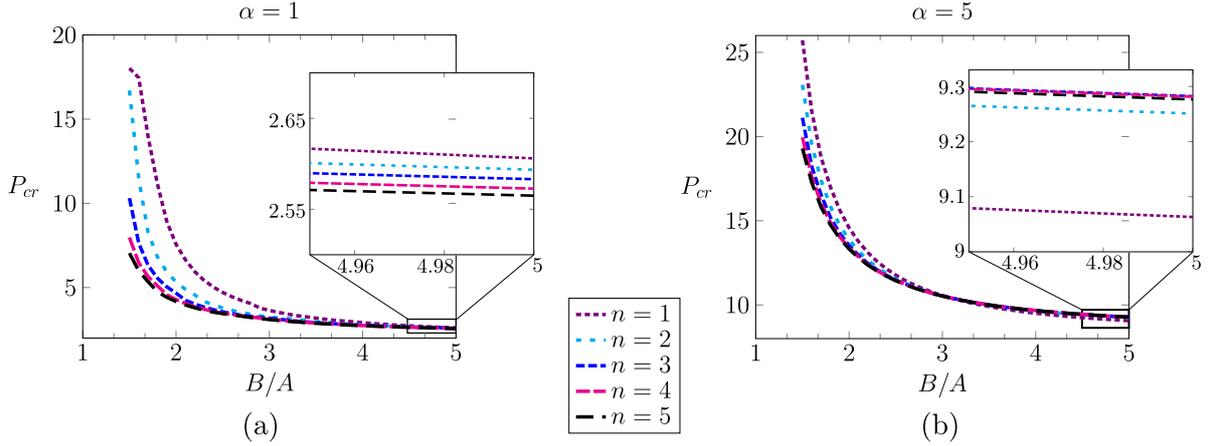}
\caption{Dependence of the critical dimensionless pressure on the radius ratio $B/A$ for bifurcation in the circumferential direction (mode number is denoted by $n$) of a constrained cylinder at (a) $\alpha = 1$, and (b) $\alpha = 5$.} \label{var_wall_press_alpha_1_and_5_circum}
\end{figure}

\begin{figure}
\centering
\begin{tabular}{c c}
\includegraphics[width=0.45\linewidth]{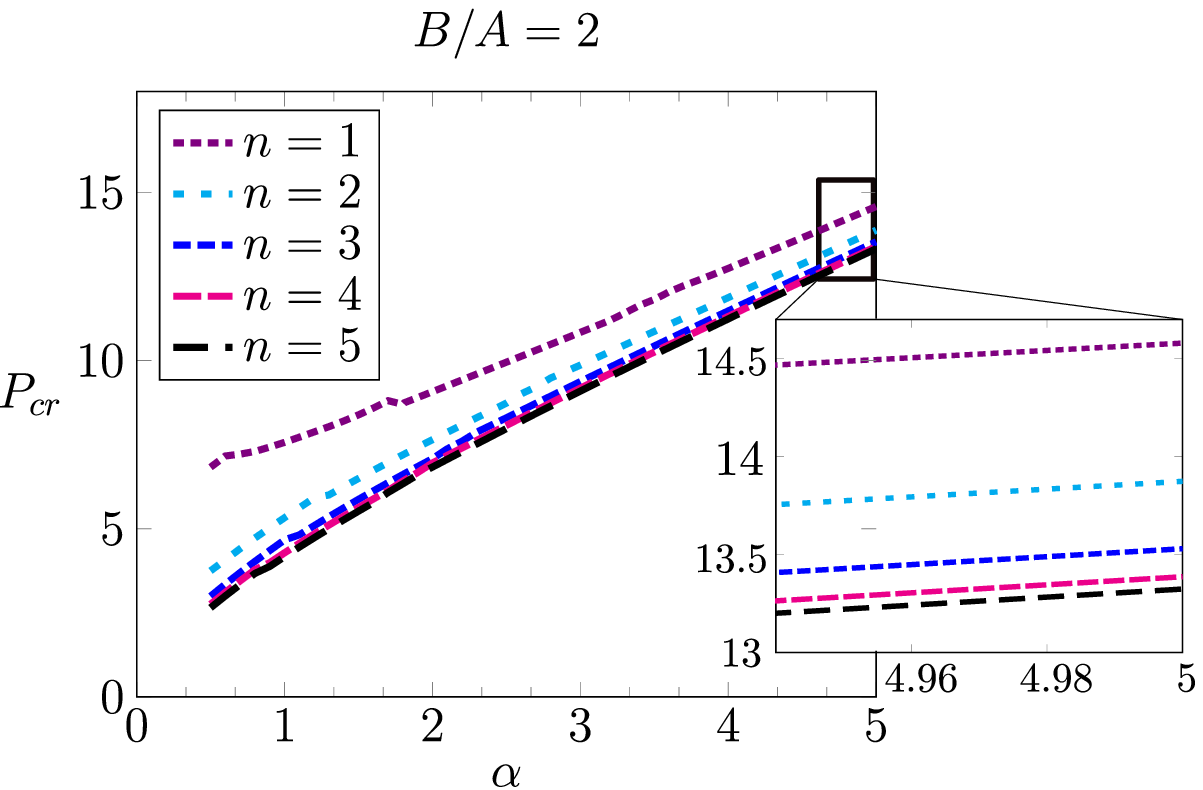}
&
\includegraphics[width=0.42\linewidth]{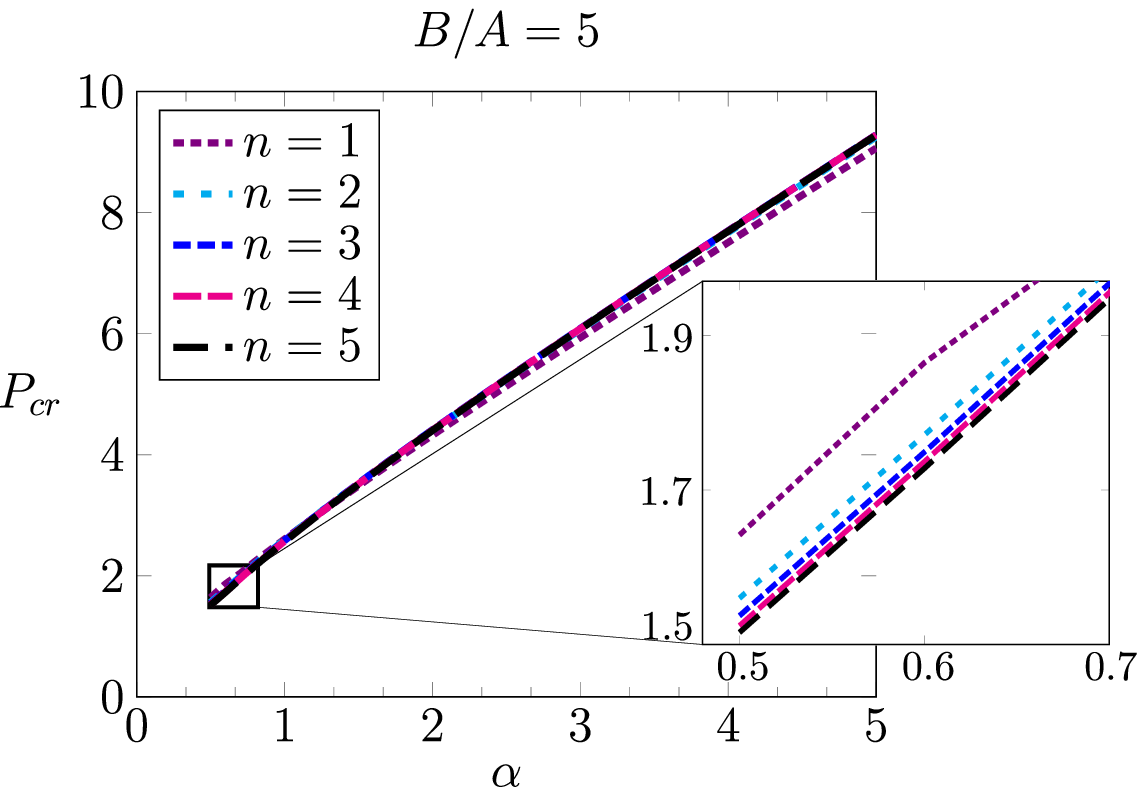}
\\
(a) & (b)
\end{tabular}
\caption{Dependence of the critical dimensionless pressure on the compressibility factor $\alpha$ for bifurcation in the circumferential direction with mode number $n$ of a constrained cylinder at (a) $B/A = 2$, and (b) $B/A = 5$.}
\label{var_alpha_press_B_A_2_and_5_circum}
\end{figure}

\begin{figure}
\centering
\includegraphics[width=0.45\linewidth]{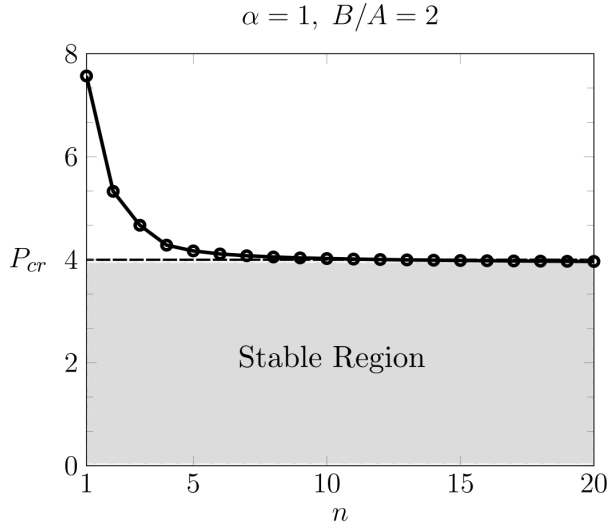}
\caption{Variation of the critical pressure for circumferential bifurcation of a constrained cylinder with respect to mode number $n$. It is seen that the curve asymptotically converges to a certain critical pressure value for all higher modes.}
\label{var_n_press_B_A_2_circum}
\end{figure}

The critical pressure to induce bifurcation in the circumferential direction is computed {numerically by solving the equations} \eqref{non_dim_first_diff_eq} and \eqref{non_dim_sec_diff_eq} subjected to the boundary conditions \eqref{non_dim_bc_1} and \eqref{non_dim_bc_2}.
Variation of the critical pressure with respect to the radius ratio $(B/A)$ {and material parameter $\alpha$ is shown in Figures \ref{var_wall_press_alpha_1_and_5_circum} and \ref{var_alpha_press_B_A_2_and_5_circum}, respectively.}
Figure \ref{var_wall_press_alpha_1_and_5_circum} shows that the critical pressure monotonically decreases with the increase in wall thickness,
but its magnitude increases with the value of $\alpha$. Due to boundary constraints,
thick cylinders withstand large deformation compared to thin cylinders and thus undergo instability at a lower critical pressure than thin cylinders. For $\alpha = 1$, the bifurcation solution of $n = 1$ requires higher pressure than the other modes suggesting that a bifurcation with higher mode number is energetically preferred to induce the instability. 
Figure \ref{var_wall_press_alpha_1_and_5_circum}b ($\alpha = 5$) shows the critical pressure curves for all modes converge earlier than the results of $\alpha = 1$.
Figure \ref{var_alpha_press_B_A_2_and_5_circum} shows the variation of critical pressure with respect to $\alpha$ for a fixed wall-thickness. 
For the thin cylinder case ($B/A = 2$), the higher modes are energetically preferred as $\alpha$ is increased. For the thick cylinder case ($B/A= 5$), the first mode $n=1$ is preferred 
with an increase in $\alpha$ value.  
The stable region for all the modes with $B/A = 2$, and $\alpha = 1$ is shown in Figure \ref{var_n_press_B_A_2_circum} which indicates the absence of bifurcation below the critical pressure, $P_{cr} \approx 4$.  

\begin{figure}
\centering
\includegraphics[width = 1\linewidth]{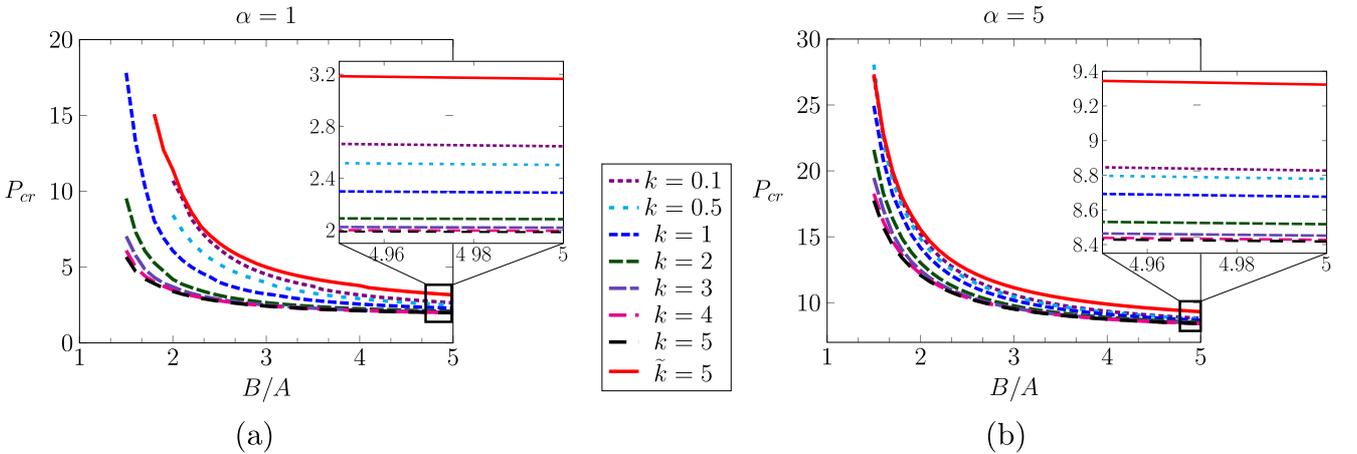}
\caption{Dependence of the critical dimensionless pressure on the radius ratio $B/A$ for bifurcation in the axial direction of a constrained cylinder at (a) $\alpha = 1$, and (b) $\alpha = 5$. The pressure curve associated with $k$ is obtained by perturbing the principal solution in radial as well as axial component of a constrained cylinder whereas the red solid pressure curve corresponds $\widetilde{k} = 5$ is obtained by perturbing the primary solution only along the radial component of the constrained cylinder.}  \label{wall_thick_vs_press_axial_constrained}
\end{figure}

\begin{figure}
\centering
\begin{tabular}{c c}
\includegraphics[width = 0.45\linewidth]{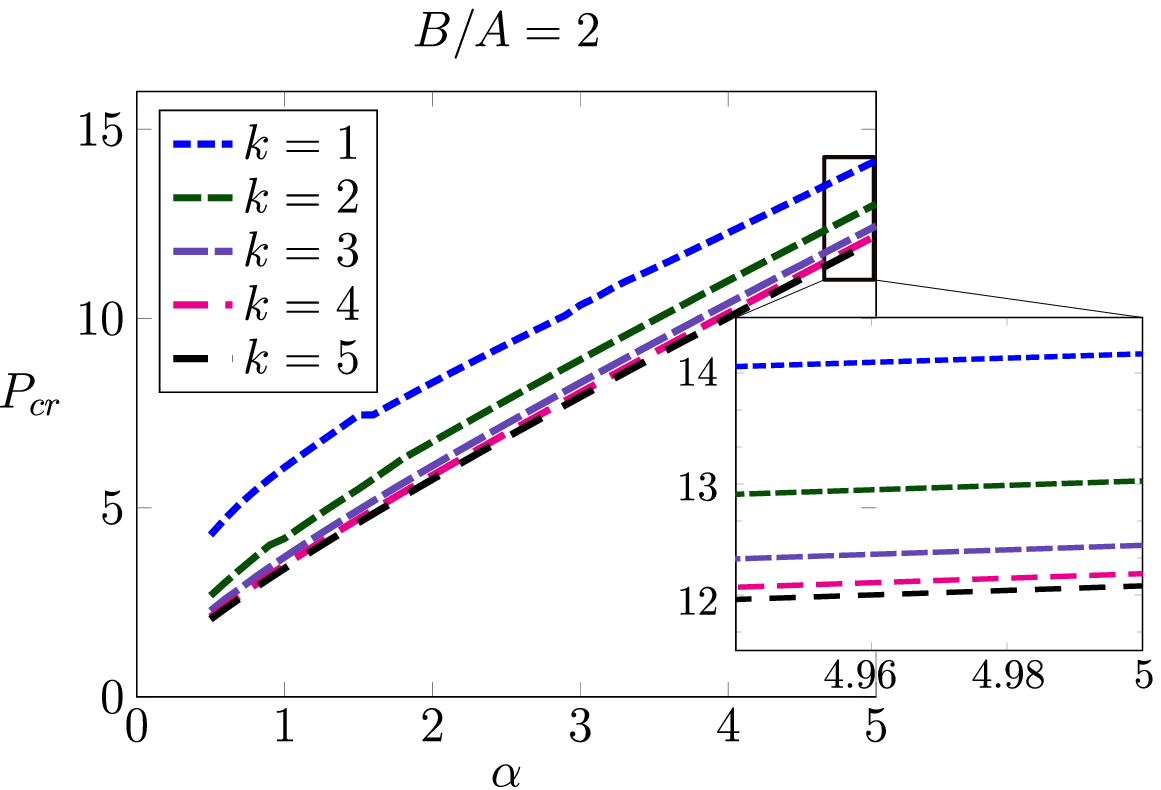}
&
\includegraphics[width = 0.45\linewidth]{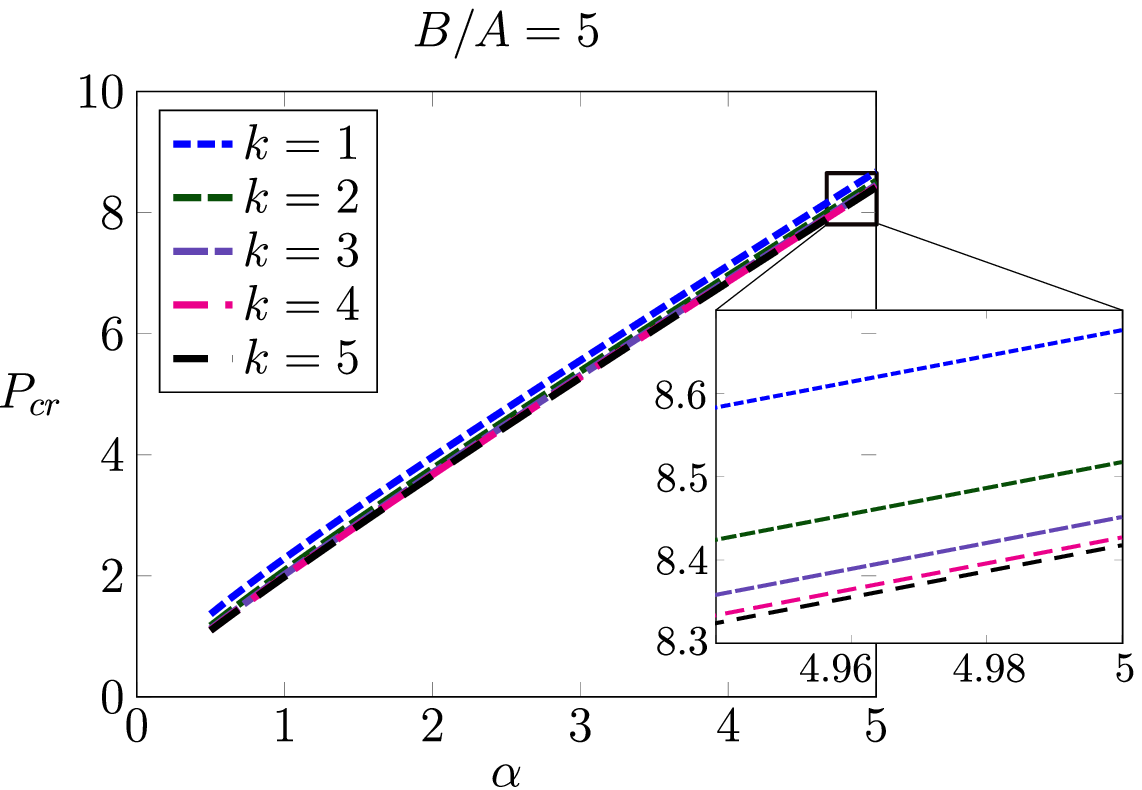}
\\
(a) & (b)
\end{tabular}
\caption{Dependence of the critical dimensionless pressure on the compressibility factor $\alpha$ for bifurcation in the axial direction of a constrained cylinder at (a) $B/A = 2$, and (b) $B/A = 5$. The pressure curve associated with $k = 5$ attains a lower bound.} \label{var_alpha_vs_press_axial_constrained}
\end{figure}

The critical pressure to induce bifurcation in the axial direction is computed by the numerical solution of equations \eqref{diff_eq_non_dim_axial_const_bound} subjected to the boundary conditions \eqref{bc_const_axial}.
Here, `$k$' is a dimensionless number which is a defined as $ k = m (2\pi/L) B$ (see Appendix \ref{reform_eqn_app}) and can be any positive number as opposed to $n$ that needs to be an integer. Higher value of $k$ corresponds to higher wavenumber ($m$) in axial direction.
Variation of the critical pressure with the radius ratio $(B/A)$ is shown in Figure \ref{wall_thick_vs_press_axial_constrained} and against the material parameter $\alpha$ in Figure \ref{var_alpha_vs_press_axial_constrained}.
The red solid pressure curve in Figure \ref{wall_thick_vs_press_axial_constrained} corresponds to the lowest critical pressure obtained by numerical solution of \eqref{one_dim_axial_const_eq} subjected to the boundary conditions \eqref{one_dim_inner_bound_const} and \eqref{one_dim_const_bound}.
This bifurcation is obtained for the mode $\widetilde{k} = 5$ and parameter values $\alpha = 1,5$.
The non-dimensional number $\widetilde{k}$ is a rescaled parameter defined as $\widetilde{k} = \widetilde{m} (2 \pi/L) B$. 
In this case, only an incremental radial strain is induced by  the critical pressure which results in the bifurcation solution that corresponds to $\widetilde{k} = 5$. 
This critical pressure is much higher than the pressure obtained for the case when  both the radial and axial strain resist the critical pressure. Thus, for $\alpha = 1,~5$, the bifurcation solution corresponds to $k = 5$ is energetically preferred over the bifurcation solution of $\widetilde{k} = 5$ to induce the instability in axial direction. The associated mathematical equations are provided in Appendix \ref{reform_eqn_app}.

The variation of critical pressure with $\alpha$, $B/A$ and $k$ is similar to that seen for the circumferential bifurcation case. 
However, the magnitude of the critical pressure obtained is smaller for all the values of the parameters chosen.
The pressure curves in Figure \ref{wall_thick_vs_press_axial_constrained} converge at higher value of $B/A > 5$ as compared to Figure \ref{var_wall_press_alpha_1_and_5_circum}. 
For the same combination of parameters $B/A = 2$ and $\alpha =1$, the pressure curves converge to limiting pressure $P_{cr} \approx 3.3$ when plotted against $k$.
Thus, a cylinder with constrained boundary subjected to an internal pressure is likely to develop instabilities with perturbations along the axial direction.  

\subsection{Bifurcation of solution for cylinder with a free external boundary}

\begin{figure}
\centering
\includegraphics[width=0.9\linewidth]{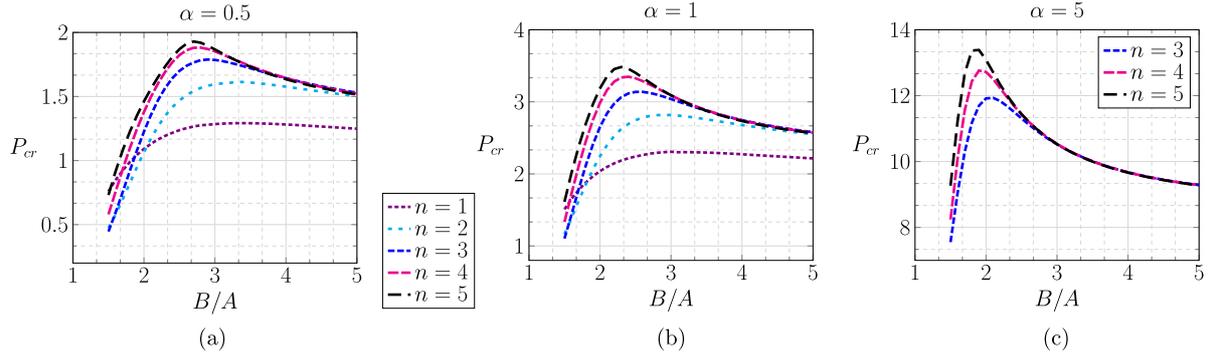}
\caption{Variation of the critical pressure against radius ratio $B/A$ for circumferential bifurcation of a cylinder with unconstrained boundary at $\alpha=0.5,~ 1~\text{and}~5$.
The lowest mode for $\alpha =5 $ occurs at $n=3$.} \label{circum_cylinder_with_free_bound}
\end{figure}

The critical pressure to induce bifurcation in the circumferential direction is computed numerically by solving equations \eqref{non_dim_first_diff_eq} and \eqref{non_dim_sec_diff_eq} subjected to stress free boundary conditions \eqref{non_dim_bc_free_case} and \eqref{non_dim_bc2_free_case}.
Variation of the critical pressure with respect to the radius ratio $(B/A)$ is shown in Figure \ref{circum_cylinder_with_free_bound}. 

The behaviour in this case is markedly different from the constrained cylinder case.
The critical pressure first rises, reaches a maximum, and then falls upon increasing the $B/A$ ratio for all the modes considered that leads to bifurcation in thick cylinders. For higher $\alpha$, the stiffness of cylindrical tube increases which results in higher extrema of critical pressure.
For $\alpha = 0.5,~1 $, the mode $n = 1$ requires less energy to induce instability compared to other modes except for lesser $B/A$ ratios as evident from the Figure \ref{circum_cylinder_with_free_bound}. Also, for $\alpha =5$, no solutions are obtained for $n=1,2$ and instability appears only for $n \geq 3$.
The value of critical pressure for all the modes with $n>1$ converge as $B/A$ increases.

\begin{figure}
\centering
\includegraphics[width= 0.9\linewidth]{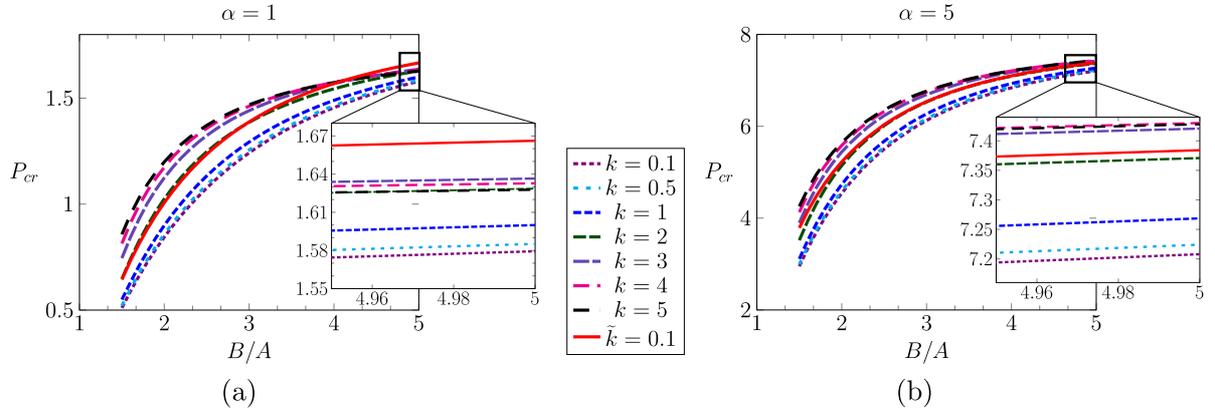}
\caption{Critical pressure variation against radius ratio for axially perturbed unconstrained cylinder at a) $\alpha = 1$, and b) $\alpha = 5$. The pressure curves associated with $k$ is obtained by perturbing the principal solution in radial-axial component of unconstrained cylinder whereas the red solid pressure curve corresponds $\widetilde{k} = 0.1$ is obtained by perturbing the primary solution only along the radial component of the unconstrained cylinder.}\label{var_wall_thick_vs_press_axial_free}
\end{figure}

\begin{figure}
\centering
\begin{tabular}{c c}
\includegraphics[width=0.5\linewidth]{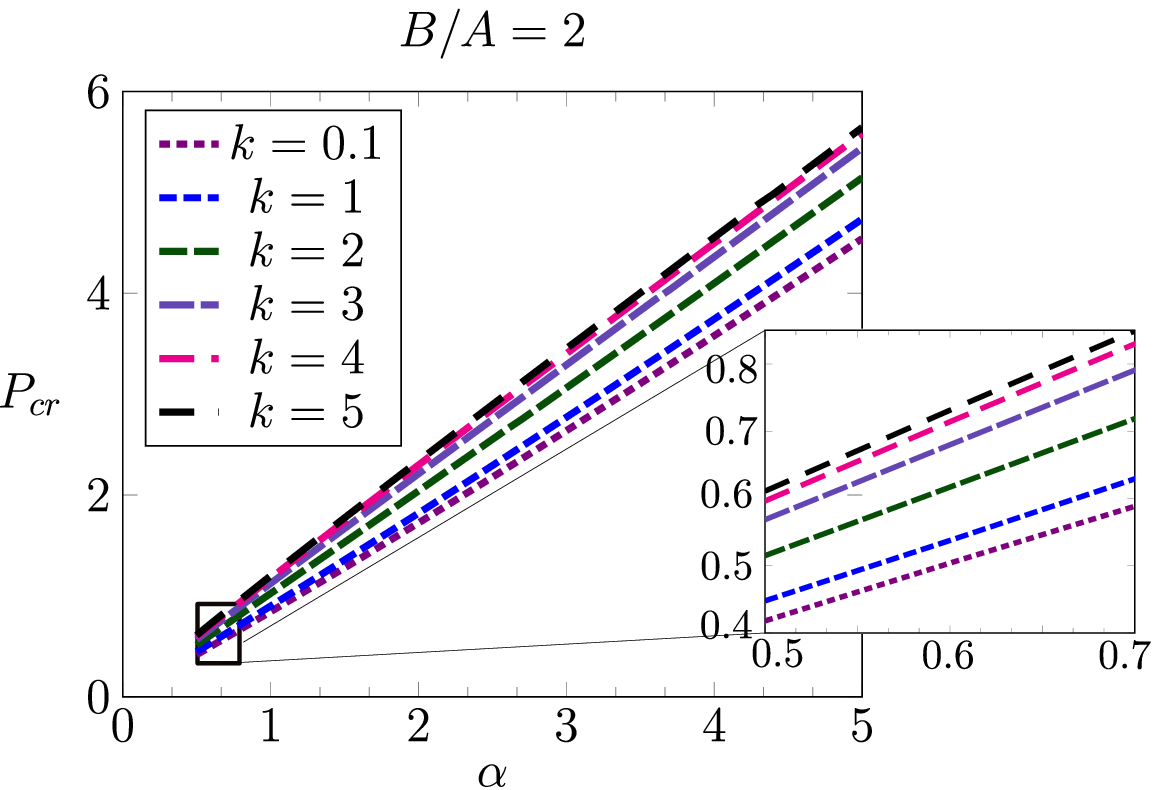}
&
\includegraphics[width=0.48\linewidth]{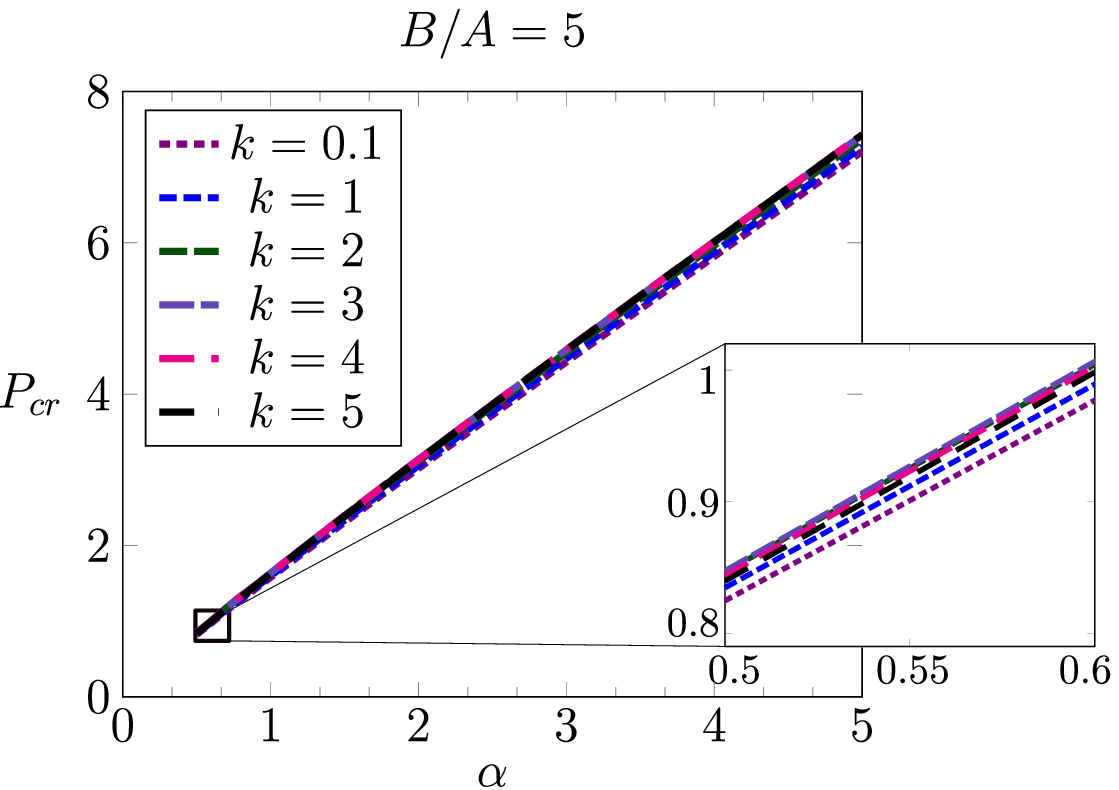}
\\
(a) & (b)
\end{tabular}
\caption{Critical pressure variation against compressibility factor for axially perturbed unconstrained cylinder at a) $ B/A = 2$, and b) $B/A = 5$. The pressure curve associated with $k = 0.1$ attains a lower bound.} \label{alpha_vs_press_axial_free}
\end{figure}

The critical pressure to induce bifurcation in the axial direction is computed by the numerical solution of equations \eqref{diff_eq_non_dim_axial_const_bound} along with the boundary conditions  \eqref{bc_const_axial_a} and \eqref{non_dim_bc_axial_free}.
Variation of the critical pressure verses radius ratio $(B/A)$ and material parameter $\alpha$ are shown in Figure \ref{var_wall_thick_vs_press_axial_free} and Figure \ref{alpha_vs_press_axial_free}, respectively.
The variation of critical pressure with $B/A$ is opposite to that observed in the case of a constrained cylinder.
$P_{cr}$ increases nonlinearly with increase in the ratio $B/A$ and all modes converge at higher $B/A$ ratios. 
Again, we have shown the onset of axial instability by perturbing only the radial component using \eqref{ansatz3_one_dim}. 
The red solid pressure curve in Figure \ref{var_wall_thick_vs_press_axial_free} is the lowest critical pressure curve obtained by the numerical solution of \eqref{one_dim_axial_const_eq} subjected to unconstrained boundary condition \eqref{one_dim_inner_bound_const} and \eqref{one_dim_free_bound} and corresponds to the mode number $\widetilde{k} = 0.1$. 
This pressure is much higher due to resistance only from the radial strain as compared to the bifurcation solution of $k = 0.1$ for $\alpha = 1,5$.
Here, the lowest wavenumber corresponds to $k=0.1$ is energetically preferred over the other modes for inducing instability along the axial direction of the cylinder.
This shows that thick cylinders have more stable behaviour at high inflation pressure and attain wrinkled configuration at a higher value of critical pressure due to large material resistance as compared to thin cylinders.
Furthermore, bifurcation for lower modes along axial direction requires less critical pressure than that for circumferential direction suggesting that buckling in axial direction is energetically preferred. The curves for $k=0.1$ and $k=0.5$ almost coincide with each other and therefore we have not shown the results for lower values of $k$.
Similar trends for threshold pressure with wave length and material stiffness ($\mu$) for incompressible cylinder with unconstrained boundary are reported by \cite{cheewaruangroj2019peristaltic}.
Similar to the constrained cylinder case, increasing the value of $\alpha$ leads to an increase in the value of the critical pressure as seen in Figure \ref{alpha_vs_press_axial_free}.
 
\subsection{Comparison of the bifurcation in the axial and circumferential directions} \label{result_comp_circum_axial}

For both the constrained and free cylinders 
, it is observed that the critical bifurcation pressure in the axial direction is lower than the circumferential direction. This can be seen by comparing the pressure curves corresponding to the lowest wavenumber in axial direction ($k=5$) in Figure \ref{wall_thick_vs_press_axial_constrained} is always lower than the pressure curve corresponds to lowest wavenumber in circumferential direction ($n = 5$)  in Figure \ref{var_wall_press_alpha_1_and_5_circum} for a constrained case. 
This same trend can be seen in unconstrained cylinder for the pressure curves corresponds to lowest wavenumber ($k = 0.1$) in axial direction in Figure \ref{var_wall_thick_vs_press_axial_free} and the pressure curve corresponds to $n = 1$ in circumferential direction in Figure  \ref{circum_cylinder_with_free_bound}.
Therefore, for a hollow cylinder made of isotropic compressible hyperelastic material, bifurcation always occurs in the axial direction as it require less pressure compared to the circumferential direction.
In order to design cylindrical systems that can lead to pattern formation (bifurcation) upon inflation in the circumferential direction, one needs to increase the stiffness in the axial direction as shown below.

\subsubsection{Stiffening of the axial direction}
Consider the cylinder to be made of an anisotropic (transversely isotropic) material with an additional stiffness along a vector $\mbf{a}$ in the reference configuration (for example, by introduction of continuously distributed fibres orientated along the vector $\mbf{a}$).
For this case, we use the elastic strain energy density function \citep{holzapfel2010constitutive}
\begin{align}
 \Omega^*(I_1,I_3,I_4) = \frac{\mu}{2} \big[ I_1 - 3 -  \, \text{log}\, I_3 \big] + \frac{\kappa}{4} \big[ \text{log}\, I_3 \big]^2  + \Omega_{f} (I_4),
\end{align}
where $\Omega_f (I_4)= \displaystyle \frac{k_1}{2 k_2} \bigg[ \exp[k_2 [I_4 - 1]^2] - 1 \bigg]$ is the energy due to fibre reinforcement, $k_1>0$ is a parameter with units of stress, and $k_2>0$ is dimensionless parameter. 
The invariant $I_4 = \mbf{a} \cdot \mbf{C}\mbf{a}$ represents the square of stretch in direction of anisotropy. 
In this case, the incremental dimensionless first  Piola--Kirchhoff stress tensor is obtained as 
\begin{align}
\frac{\delta \mathbf{{P}}}{\kappa} = &  \alpha  \bigg[ \delta \mathbf{F} + {[\mathbf{F}^{-1} [\delta \mathbf{F}]~ \mathbf{F}^{-1}]^{T}} \bigg] + 2 \bigg[ \mathbf{F}^{-T}~ \text{tr}(\mbf{F}^{-1} [\delta \mathbf{F}]) - \log J {[\mathbf{F}^{-1} [\delta \mathbf{F}]~ \mathbf{F}^{-1}]^{T}} \bigg] \nonumber\\
& +  \Bigg[2 \bar{k}_1 \exp \big[k_2 [I_4 - 1]^2 \big] \bigg[ 1 + 2 k_2 [I_4 - 1] \bigg] \big[\mbf{a} \otimes \mbf{Fa} \big] \bigg[ \text{tr} \bigg( \delta \mbf{F}^T [\mbf{a} \otimes \mbf{Fa}] \bigg)\bigg] \nonumber\\
& + 2 \bar{k}_1 [I_4 - 1] \exp \big[k_2 [I_4 - 1]^2 \big] \left[\mbf{a} \otimes \mbf{a} \right] \text{tr}(\delta \mbf{F}^T), \label{inc_stress_fibers}
\end{align}
where $\bar{k}_1 = k_1/\kappa$ is a dimensionless parameter. 
In our problem we assume plane strain and that the anisotropy is orientated along the axis of the cylinder that results in  $I_4 = \lambda_Z^2 =1$  and we obtain
\begin{align}
\frac{\delta \mathbf{P}}{\kappa} = & \alpha \bigg[ \delta \mathbf{F} + {[\mathbf{F}^{-1} [\delta \mathbf{F}]~ \mathbf{F}^{-1}]^{T}} \bigg] + 2 \bigg[ \mathbf{F}^{-T}~ \text{tr}[\mbf{F}^{-1} [\delta \mathbf{F}]] - \log J {[\mathbf{F}^{-1} [\delta \mathbf{F}]~ \mathbf{F}^{-1}]^{T}} \bigg] \nonumber\\
& \hspace{2.5in} + 2 \bar{k}_1 \big[\mbf{a} \otimes \mbf{Fa} \big] \bigg[ \text{tr} \bigg( \delta \mbf{F}^T [\mbf{a} \otimes \mbf{Fa}] \bigg)\bigg].
\end{align}
Auxiliary calculations to arrive at the above equations are provided in Appendix \ref{Fibers_app}.

\begin{figure}
\centering
\begin{tabular}{c}
\begin{tabular}{c c}
\includegraphics[width = 0.44\linewidth]{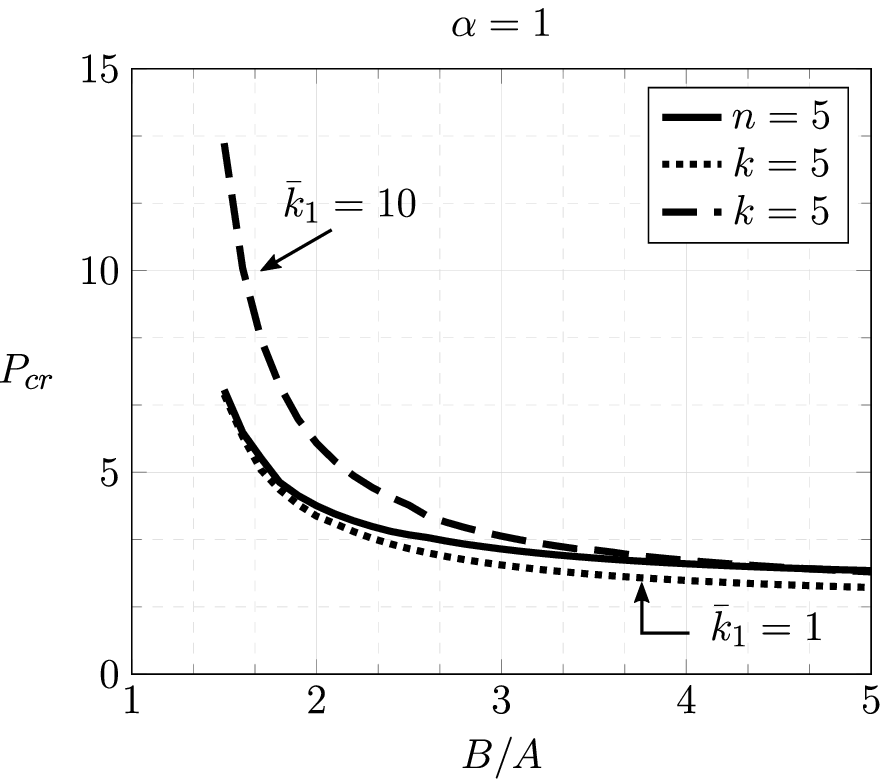}
&
\includegraphics[width = 0.44\linewidth]{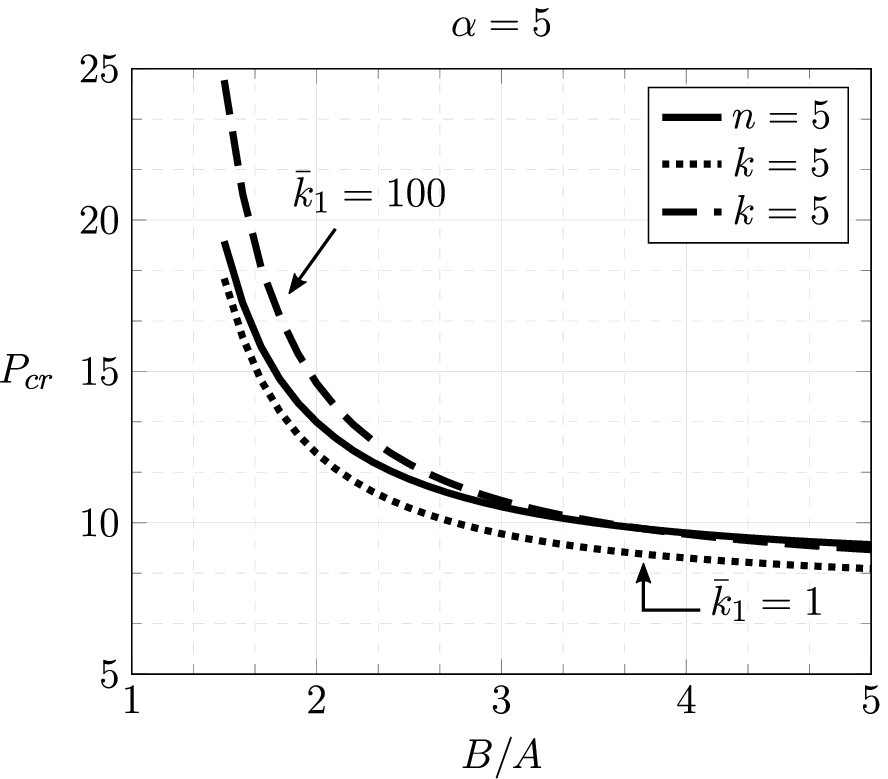}
\\
(a) & (b)
\end{tabular}
\\
\includegraphics[width = 0.44\linewidth]{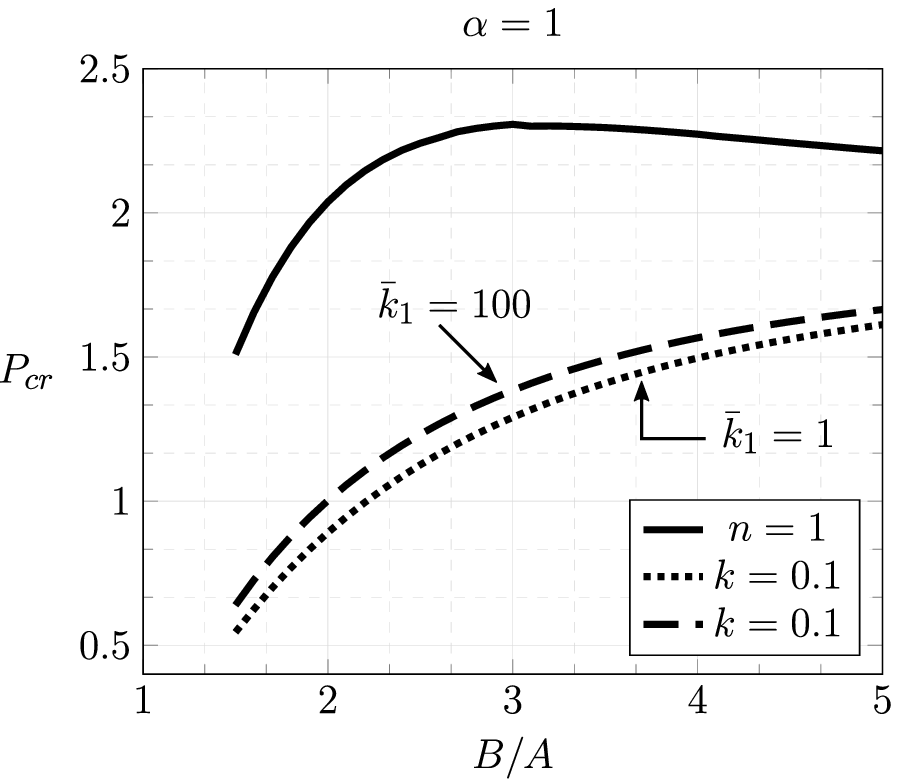}
\\
(c)
\end{tabular}
\caption{Variation of the critical pressure against radius ratio for a cylinder reinforced along the axial direction. The solid curve corresponds to bifurcation in the circumferential direction and the dashed curves corresponds to bifurcation in the axial direction for cylinder with a displacement constraint on the external boundary, (a) $\alpha = 1$, (b) $\alpha = 5$ and (c) free external surface with $\alpha = 1$.} 
\label{fibers_axial}
\end{figure}

The stress at the material point not only depend on the deformation gradient $\mbf{F}$ but also the fibre direction $\mbf{a}$. For, the cylinder with unit axial stretch ($\lambda_Z = 1$), the incremental stress corresponding to fibre term is independent of the dimensionless parameter $k_2$.
The influence of stiffening along the axial coordinate on the critical pressure is demonstrated in Figure \ref{fibers_axial}. The plots show the variation of critical pressure with thickness for the lowest wave numbers ($n = 5$) for circumferential and $ k = 5$ for axially perturbed cylinder. Figure \ref{fibers_axial}a (respectively, \ref{fibers_axial}b) corresponds to cylinder with constrained outer surface for $\alpha = 1$ (respectively, $\alpha = 5$).  As the stiffness value $\bar{k}_1$ is increased, the critical pressure required to achieve bifurcation along the axial direction increases significantly  for $B/A \leq 3$ in constrained cylinder. The stiffness value also depends on the material parameter, higher value of $\alpha$ requires higher stiffening as shown in Figure \ref{fibers_axial}b.
This makes the bifurcation along the circumferential direction more preferable and provides a mechanism for tuning the bifurcation characteristics of such systems.  However, in unconstrained cylinder, the stiffening along axial coordinate has no significant effect on critical pressure even for very high stiffening value ($\bar{k}_1  = 100$) as shown in Figure \ref{fibers_axial}c. The bifurcation always occurs at lower value of critical pressure in axial direction when compared to circumferential direction for unconstrained cylinder as discussed in the previous Section \ref{result_comp_circum_axial}.

\section{Conclusion} \label{section_conclusion}

In summary, we have studied large deformation in internally pressurised thick-walled compressible cylinders made up of soft material due to their widespread applications in biomedical implants, additively manufactured metamaterials, highly flexible/stretchable electronics, soft microfluidic channels and soft robotics.  

The extreme internal pressure leads to elastic instability in thick-walled compressible cylindrical channels along the  circumferential or axial direction. 
Incremental deformation theory is applied to derive the governing PDEs for these cylindrical channels.
Two types of boundary conditions for the external surface of the cylinder are studied, namely, constrained and unconstrained to comprehend bifurcation phenomenon in the circumferential or axial direction.
The resulting incremental differential equations are obtained by perturbing the primary solution along the radial-circumferential as well as radial-axial direction. 
These equations are numerically solved for both the boundary conditions using compound matrix method and shooting method to obtain the critical internal pressure which induces the instability.
We have also investigated the elastic instability in axial direction by perturbing the primary solution only along the radial component of the cylinder. This results in higher critical pressure as compared to the critical pressure obtained through generalised radial-axial perturbation for both the boundaries considered.
The effect of radius ratio (wall-thickness), compressibility factor and boundary conditions on the critical inflating pressure are systematically studied.

We also demonstrate that the numerical solutions of the resulting ODEs can be computed almost three times faster using compound matrix method as compared to simple shooting method.
We observe that the pressure curves associated with constrained external surface have shown opposite behaviour than stress free external surface. The critical pressure decreases with the increase of radius ratio due to the fixed boundary conditions in the constrained cylinder whereas the critical pressure increases with the radius ratio in the unconstrained boundary condition case.
For constrained cylinder, the pressure curves asymptotically converges with the increase of wavenumber, therefore bifurcation solution corresponds to higher wavenumber is energetically preferred. The explicit value of critical pressure is difficult to obtain, thus the stable region for the optimised working pressure is provided in which bifurcation is absent. 
Our computations reveal that for the lowest stable mode, the critical pressure that causes bifurcation in the axial direction is always lower than the critical pressure that causes bifurcation in the circumferential direction. However, this observation does not hold when the axial direction is stiffened with the fibres. The reinforcement of fibres in axial direction causes the bifurcation along the circumferential direction is more preferable in constrained cylinders whereas reinforcement have very less effect on bifurcation solution for unconstrained cylinders. 
We have restricted ourselves to determine the threshold pressure, but a post-bifurcation analysis may provide insights on the amplitude of wrinkles and stability of wrinkled solution. These avenues are currently under investigation.

 \section*{Acknowledgement} 
 Prashant Saxena acknowledges the support of startup funds from the James Watt School of Engineering at the University of Glasgow.
 The authors thank Prof Ray W Ogden for his valuable suggestions to improve the manuscript.

\bibliographystyle{author}
\bibliography{annulus2}

\newpage
\begin{appendix}
\section{Appendix: Incremental stress and  traction condition} \label{Inc_stress_app}
The incremental stress \eqref{Incremental_stress} is rewritten in index notation as
\begin{align}
[\delta {\mbf{P}}]_{ij}&= \mathcal{A}_{ijkl}^{(1)}~ [\delta {\mbf{F}}]_{kl}, \label{incr_piola}
\end{align}
where $\mathcal{A}^{(1)}_{ijkl}$ is the first order elastic moduli given by 
\begin{equation}
\begin{aligned}
\mcal{A}_{ijkl}^{(1)} & = \mu \bigg[ \delta_{ik} \delta_{jl} - \big[\mathbb{T}[-\mathbf{F}^{-1} \boxtimes \mathbf{F}^{-T}] \big]_{ijkl} \bigg] + 2 \kappa \bigg[ [\mathbf{F}^{-T}]_{ij} [\mathbf{F}^{-T}]_{kl} + \log J~ \left[\mathbb{T}[-\mathbf{F}^{-1} \boxtimes \mathbf{F}^{-T}]\right]_{ijkl}\bigg],\\
&=\mu\bigg[\delta_{ik} \delta_{jl} + F_{jk}^{-1} F_{il}^{-T} \bigg]+ 2 \kappa \bigg[  F_{ij}^{-T} F_{kl}^{-T} - \log J [F_{jk}^{-1} F_{il}^{-T}] \bigg].
\end{aligned}
\end{equation}
Using \eqref{elasti_modulii}, the Piola Kirchhoff stress is obtain as
\begin{align}
[\delta {\mbf{P}}]_{ij} & = \Bigg[\mu \bigg[\delta_{ik} \delta_{jl} +  F_{jk}^{-1} F_{il}^{-T} \bigg] + 2 \kappa \bigg[ F_{ij}^{-T} F_{kl}^{-T} - \log J [F_{jk}^{-1} F_{il}^{-T}] \bigg] \Bigg] [\delta \mbf{F}]_{kl}, \nonumber \\[10pt]
& = \mu \bigg[ [\delta \mbf{F}]_{ij} + [\mbf{F}^{-1} [\delta \mbf{F}] \mbf{F}^{-1}]_{ji}\bigg] + 2 \kappa \bigg[F_{ij}^{-T}~ [{F}^{-T}_{kl} [\delta \mbf{F}]_{kl}] - \log J [\mbf{F}^{-1} [\delta \mbf{F}] \mbf{F}^{-1}]_{ji} \bigg], \nonumber \\[10pt]
& = \mu \bigg[ [\delta \mbf{F}]_{ij} + {[\mbf{F}^{-1} [\delta \mbf{F}] \mbf{F}^{-1}]^{T}}_{ij} \bigg] \nonumber \\
& \qquad + 2 \kappa \bigg[F_{ij}^{-T}~ [\mbf{F}^{-1} [\delta \mbf{F}]]_{kk} - \log J {[\mbf{F}^{-1} [\delta \mbf{F}] \mbf{F}^{-1}]^{T}}_{ij}\bigg]. \label{inc_stress_index}
\end{align}
Using \eqref{inc_stress_index}, the Piola stress in direct notation is  
\begin{align}
\delta \mathbf{P} = \mu \bigg[ \delta \mathbf{F} + {[\mathbf{F}^{-1} [\delta \mathbf{F}]~ \mathbf{F}^{-1}]^{T}} \bigg] + 2 \kappa \bigg[ \mathbf{F}^{-T}~ \text{tr}[\mbf{F}^{-1} [\delta \mathbf{F}]] - \log J {[\mathbf{F}^{-1} [\delta \mathbf{F}]~ \mathbf{F}^{-1}]^{T}} \bigg]. 
\end{align}
 Further, using \eqref{trans_cauchy_to_Piola}, 
the incremental traction condition for inflating cylinder is given as
\begin{align}
[\mbf{P} + \delta \mbf{P}] \mbf{N}& = -\bigg[J + \frac{\partial J}{\partial \mbf{F}} \cdot \delta \mbf{F} \bigg] \bigg[ P_r + d P_r\bigg] \bigg[\mbf{F}^{-T} + \frac{\partial \mbf{F}^{-T}}{\partial \mbf{F}}\cdot \delta \mbf{F} \bigg] \mbf{N}, \nonumber\\
& = - \bigg[J P_r + J dP_r + P_r  \bigg[\frac{\partial J}{\partial \mbf{F}} \cdot \delta \mbf{F} \bigg] \bigg] \bigg[\mbf{F}^{-T} + \frac{\partial \mbf{F}^{-T}}{\partial \mbf{F}}\cdot \delta \mbf{F} \bigg] \mbf{N}, \nonumber \\
& = - \bigg[J P_r \mbf{F}^{-T} \mbf{N} + J dP_r \mbf{F}^{-T} \mbf{N} + J P_r \bigg[\frac{\partial \mbf{F}^{-T}}{\partial \mbf{F}}\cdot \delta \mbf{F} \bigg] \mbf{N} +  P_r  \bigg[\frac{\partial J}{\partial \mbf{F}} \cdot \delta \mbf{F} \bigg] \mbf{F}^{-T} \mbf{N} \bigg].
\end{align}
This results in 
\begin{align} 
[\delta \mbf{P}] \mbf{N} &= -J P_r \bigg[ -\mathbb{T}\big[\mbf{F}^{-1} \boxtimes \mbf{F}^{-T} \big] \cdot [\delta \mbf{F}] \bigg] \mbf{N} - J dP_r \mbf{F}^{-T} \mbf{N} - P_r \big[ \det(\mbf{F}) \mbf{F}^{-T} \cdot [\delta \mbf{F}] \big] \mbf{F}^{-T} \mbf{N}, \nonumber\\
&= -J P_r \bigg[ - \big[F^{-1}_{jk} F^{-T}_{il} \big] [\delta \mbf{F}]_{kl} \bigg] N_j- J dP_r F^{-T}_{ij} N_j - J P_r ~ \text{tr}(\mbf{F}^{-1} [\delta \mbf{F}]) F_{ij}^{-T} N_j, \nonumber \\
&=J P_r \big[ F^{-T}_{il}~ [\delta \mbf{F}]^{T}_{lk}~ F^{-T}_{kj} \big] N_j - J dP_r F^{-T}_{ij} N_j - J P_r ~ \text{tr}(\mbf{F}^{-1} [\delta \mbf{F}]) F_{ij}^{-T} N_j. \label{del_P_N}
\end{align}
Eq. \eqref{del_P_N} can be written in direct notation as
\begin{align}
[\delta \mbf{P}]\mbf{N}=& J P_r \mbf{F}^{-T} ~[\delta \mbf{F}]^{T} ~\mbf{F}^{-T} \mbf{N} - J ~ dP_r \mbf{F}^{-T} \mbf{N} - J P_r ~ \text{tr}(\mbf{F}^{-1} [\delta \mbf{F}]) \mbf{F} \mbf{N}.
\end{align}
\section{Reformulation of equations and numerical solution}\label{reform_eqn_app}

\subsection{Case 1: Circumferential perturbations with constrained boundary}\label{subsec_circum_const}

In order to perform efficient numerical computations, we define the dimensionless parameters 
\begin{align}
\rho = \frac{R}{B}, \quad \rho_1 = \frac{r}{B}, \quad f = \frac{\Delta f}{B}, \quad g = \Delta g, \label{non_dim}
\end{align} 
where $B$ is the outer radius of constrained cylinder. On substitution of \eqref{non_dim} in the governing equations \eqref{diff_after_comp_eps_1} and \eqref{diff_after_comp_eps_2}, we obtain the incremental differential equations in terms of dimensionless displacements $f$ and $g$  as
\begin{align}
f'' & = - \frac{1}{a_1}\bigg[a_2 f' + a_3 f +a_4 g' + a_5 g \bigg],\label{non_dim_first_diff_eq} \\
g''&=-\frac{1}{b_1} \bigg[b_2 g' + b_3 g + b_4 f'  + b_5 y_1 \bigg],\label{non_dim_sec_diff_eq}
\end{align}
where
\begin{align*}
a_1 &=  \rho_1' \rho_1^2 \rho^2 \bigg[\alpha \rho_1'^2 - 2  \log \left( \frac{\rho_1 \rho_1'}{\rho}\right) + 2  + \alpha \bigg] f'', \\
a_2 & =  \rho_1 \rho \bigg[ \big[2 \rho_1'' \rho_1 \rho - \rho_1' \rho_1 \big] 2 \log \left( \frac{\rho_1 \rho_1'}{\rho}\right) + \rho_1'^3 \rho_1 \alpha - 6 \rho_1'' \rho_1 \rho  - 2 \rho_1'' \rho_1 \rho \alpha - 2 \rho_1'^2 \rho + 4 \rho_1' \rho_1  + \rho_1' \rho_1 \alpha \bigg],  \\
a_3 &=  \rho_1' \bigg[ -\rho_1'^2 \rho_1^2 \alpha n^2 + \rho_1'^2 \rho^2 2 \log \left( \frac{\rho_1 \rho_1'}{\rho}\right) - 2 \rho_1'' \rho_1 \rho^2  - \rho_1'^2 \rho_1^2 \alpha - 4 \rho_1'^2 \rho^2 - \rho_1'^2 \rho^2 \alpha + 2 \rho_1' \rho_1 \rho \bigg],  \\
a_4 &=   -\rho_1'^2 \rho_1^2 \rho^2 n \bigg[2  \log \left( \frac{\rho_1 \rho_1'}{\rho}\right) - 2  - \alpha \bigg], \\
a_5 & = \rho_1' \rho_1 n \bigg[ 2 \log \left( \frac{\rho_1 \rho_1'}{\rho}\right)\rho_1'^2 \rho^2 - 2 \rho_1'' \rho_1 \rho^2 - \rho_1'^2 \rho_1^2 \alpha - 2 \rho_1'^2 \rho^2  - \rho_1'^2 \rho^2 \alpha + 2 \rho_1' \rho_1 \rho  \bigg], 
\end{align*}
and
\begin{align*}
b_1 & =  \bigg[\rho_1'^2 \rho_1^2 \rho^2 \alpha \bigg], \quad b_2 = - \rho_1' \rho_1 \rho \bigg[ -\rho_1'^2 \rho \alpha + 2  \log \left( \frac{\rho_1 \rho_1'}{\rho}\right) \rho - \rho_1' \rho_1 \alpha - \rho \alpha \bigg], \\
b_3 & = \rho_1'^2 n^2 \bigg[ 2  \log \left( \frac{\rho_1 \rho_1'}{\rho}\right) \rho^2 - \rho_1^2 \alpha - 2 \rho^2  - \rho^2 \alpha \bigg], \quad b_4 = \rho_1' \rho^2 n \bigg[ 2  \log \left( \frac{\rho_1 \rho_1'}{\rho}\right) - 2 -\alpha \bigg],  \\
b_5 &=  -n \bigg[ \big[ \rho_1'' \rho^2 - \rho_1' \rho \big] 2  \log \left( \frac{\rho_1 \rho_1'}{\rho}\right) - 2 \rho_1'' \rho^2   - \rho_1'' \rho^2 \alpha +2 \rho_1'^2 \rho_1 \alpha + 2 \rho_1' \rho  + \rho_1' \rho \alpha \bigg], 
\end{align*}
\noindent subjected to non-dimensionalised boundary conditions at the inner surface of cylinder (at $\rho = A/B = A^*$) 
\begin{subequations} \label{non_dim_bc_1}
\begin{align}
[2 \rho_1' A^* ] f - \rho_1 \bigg[ -\rho_1'^2 A^* \alpha +  A^* \left[ 2  \log \left( \frac{\rho_1 \rho_1'}{A^*}\right)  \right] + \rho_1 \rho_1' \wt{P} - 2 A^*  -  A^*  \alpha \bigg] f'  +  [2 \rho_1 \rho_1' A^* n] g = 0,\\
n \bigg [A^* \left[ 2 \log \left( \frac{\rho_1 \rho_1'}{A^*} \right) \right] +\rho_1 \rho_1' \wt{P} - A^* \alpha \bigg] f + [A^* \rho_1^2 \rho_1' \alpha] g'  = 0,
\end{align}
\end{subequations}
where $\wt{P}=P_r/ \kappa$ and $A^* = \left. \rho  \right|_{\text{at}{A/B}}$. 
\noindent The constrained outer boundary at $\rho=1$ leads to the condition
\begin{align}
f(1) = g(1) =0. \label{non_dim_bc_2}
\end{align}

\subsection{Case 2: Circumferential perturbations with free boundary}\label{subsec_circum_free}
Using equation \eqref{traction_bc} and the dimensionless parameters \eqref{non_dim}, the boundary condition at the inner boundary, $\rho = A^*$ for free cylinder is given by 
\begin{subequations}\label{non_dim_bc_free_case}
\begin{align}
[2 \rho_1' A^* ]f- \rho_1 \Bigg[ -\rho_1'^2 A^* \alpha +  A^* \left[ 2  \log \left( \frac{\rho_1 \rho_1'}{A^*}\right)  \right] + \rho_1 \rho_1' \wt{P} - 2 A^*  -  A^*  \alpha \Bigg] f'  +  [2 \rho_1 \rho_1' A^* n] g = 0,\\
n \bigg [A^* \left[ 2 \log \left( \frac{\rho_1 \rho_1'}{A^*} \right) \right] +\rho_1 \rho_1' \wt{P} - A^* \alpha \bigg] f + [A^* \rho_1^2 \rho_1' \alpha] g  = 0,
\end{align} 
\end{subequations}
and at the outer boundary, $\rho =  1$ is 
\begin{subequations} \label{non_dim_bc2_free_case}
\begin{align} 
[2 \rho_1' ]f - \rho_1 \Bigg[ - \rho_1'^2 \alpha +  \left[ 2  \log \left( \frac{\rho_1 \rho_1'}{1}\right)  \right]  - 2   -   \alpha \Bigg] f'  +  [2 \rho_1 \rho_1'  n] g = 0,\\
n \bigg [\left[ 2 \log \left( \frac{\rho_1 \rho_1'}{1} \right) \right] -  \alpha \bigg] f + [ \rho_1^2 \rho_1' \alpha] g  = 0.
\end{align}
\end{subequations}

\subsection{Case 3: Axial perturbations with constrained boundary}\label{subsec_axial_const}

We define the dimensionless parameters
\begin{align} 
\rho  =  \frac{R}{B}, \quad  \rho_1 = \frac{r}{B}, \quad \bar{f} = \frac{\Delta \bar{f}}{B}, \quad \bar{h} = \frac{\Delta \bar{h}}{B},  \quad k  =  m \frac{2 \pi}{L} B ,
\label{eq: dimensionless axial}
\end{align}
that lead to reformulation of the governing equations \eqref{first_order_diff_eq_axial} as
\begin{equation}\label{diff_eq_non_dim_axial_const_bound}
\begin{aligned}
{\bar{f}}''&=-\frac{1}{c_1^*}\bigg[c_2^* {\bar{f}}' + c_3^* \bar{f} +c_4^* {\bar{h}}' + c_5^* \bar{h} \bigg], \qquad {\bar{h}}''&=-\frac{1}{d_1^*} \bigg[d_2^*  {\bar{h}}' + d_3^*  \bar{h} + d_4^*  {\bar{f}}' + d_5^* {\bar{f}} \bigg],
\end{aligned}
\end{equation} 
where the dimensionless coefficients are given by
\begin{align*}
c_1^* &= {\rho_1}' {\rho_1}^2 {\rho}^2 \bigg[ {\rho_1'}^2 \alpha - 2 \log \left( \frac{\rho_1 \rho_1'}{\rho} \right) + 2 + \alpha \bigg],\\
c_2^* &= \rho_1 \rho \bigg[ {\rho_1'}^3 \rho_1 \alpha + [2 \rho_1 \rho_1'' \rho - \rho_1' \rho_1]~ 2 \log \left( \frac{\rho_1 \rho_1'}{\rho} \right) -  2 \rho_1'^2 \rho - 6 \rho_1 \rho_1'' \rho - 2 \rho_1 \rho_1'' \rho \alpha + 4 \rho_1' \rho_1 + \rho_1' \rho_1  \alpha  \bigg], \\
c_3^* & = -\rho_1' \bigg[ \rho_1'^2 \rho_1^2 \rho^2 \alpha k^2 - 2 \log \left(\frac{\rho_1 \rho_1'}{\rho}\right) \rho_1'^2 \rho^2  + {\rho_1'}^2 \rho_1^2 \alpha + 4 {\rho_1'}^2 \rho^2 + {\rho_1'}^2 \rho^2 \alpha + 2 \rho_1 \rho_1'' \rho^2 - 2 \rho_1' \rho_1 \rho \bigg],\\
c_4^* & = -\rho_1'^2 \rho_1^2 \rho^2 k \bigg[2 \log \left( \frac{\rho_1 \rho_1'}{\rho} \right) - 2 -\alpha \bigg],\qquad
c_5^* = -2 \rho_1' \rho_1 \rho k  \bigg[ \rho_1'^2 \rho + \rho_1 \rho_1'' \rho - \rho_1' \rho_1 \bigg],
\end{align*}
\begin{align*}
d_1^* &= \rho_1'^2 \rho \alpha, \qquad
d_2^* = \rho_1'^2 \alpha, \qquad
d_3^*  = k^2 \rho \rho_1'^2 \bigg[ 2 \log \left(\frac{\rho_1 \rho_1'}{\rho}\right) - 2 - 2 \alpha \bigg],\nonumber \\
d_4^* & = \rho_1' \rho k \bigg[ 2 \log \left(\frac{\rho_1 \rho_1'}{\rho}\right) - 2 - \alpha \bigg], \qquad
d_5^*  =   k \bigg[ \rho_1' - \rho_1'' \rho] ~ 2 \log \left( \frac{\rho_1 \rho_1'}{\rho}\right) \bigg] + 2 \rho_1''\rho + \rho_1'' \rho \alpha - 2 \rho_1' - \rho_1' \alpha \bigg],
\end{align*}
subjected to constrained boundary conditions 
\begin{subequations} \label{bc_const_axial}
\begin{equation}
\left.
\begin{aligned}
c_{11}^* \bar{f} + c_{22}^*  {\bar{f}}' + c_{33}^*  \bar{h} &= 0,\label{bc_const_axial_a} \\
d_{11}^*  \bar{f} + d_{44}^*  \bar{h}' &= 0, 
\end{aligned} \right\} \qquad \text{at} \quad \rho= \frac{A}{B},
\end{equation}
\begin{align}
 \bar{f}(1) =  \bar{h}(1) &= 0, \qquad  \text{at} \qquad \rho=1, \label{axial_const_bc}
\end{align}
\end{subequations}
and
\begin{align*}
c_{11}^* &=  2 \rho_1' A^*, \qquad c_{22}^* =  -\rho_1 \bigg[ -\rho_1'^2 A^* \alpha + 2 \log \left(\frac{\rho_1 \rho_1'}{A^*}\right) A^* + \rho_1 \wt{P} \rho_1' - 2 A^* - A^* \alpha \bigg], \\
\qquad c_{33}^*  & = 2 \rho_1 \rho_1' A^* k, \qquad
d_{11}^*  = k \bigg[ 2 \log \left(\frac{\rho_1 \rho_1'}{A^*}\right) A^* + \rho_1 \wt{P} \rho_1' - A^* \alpha \bigg], \qquad
d_{44}^*  = \alpha \rho_1' A^*,
\end{align*}
here  $A^* =  \left. \rho  \right|_{\text{at} A/B} $.

\subsection{Case 4: Axial perturbations with free boundary}\label{subsec_axial_free}

The incremental differential equations for cylindrical channels with unconstrained boundary is same as \eqref{diff_eq_non_dim_axial_const_bound} and the inner boundary ($\rho = A/B$) subjected to internal pressure is same as \eqref{bc_const_axial_a}. The boundary condition at the outer boundary at $\rho = 1 $ is given by 
\begin{subequations} \label{non_dim_bc_axial_free}
\begin{align} 
[2 \rho_1' ] \bar{f} -\rho_1 \bigg[ -\rho_1'^2 \alpha + 2 \log \left(\rho_1 \rho_1' \right) A^*  - 2 - \alpha \bigg] \bar{f}' + 2 \rho_1 \rho_1' k \bar{h} = 0,\\
k \bigg[ 2 \log \left(\rho_1 \rho_1' \right)  - \alpha \bigg] \bar{f} + \alpha \rho_1' \bar{h}'  = 0.
\end{align}
\end{subequations}

\subsection{Perturbation along radial component of cylinder for axial bifurcation} \label{One_dim_perturbation_app}
Upon substituting \eqref{ansatz3_one_dim} in \eqref{inc_eqbm_eq}, then rescaling the obtained equation using non-dimensional terms as $ \rho  =  \displaystyle \frac{R}{B}, \quad \rho_1 = \displaystyle \frac{r}{B}, \quad \widetilde{f} = \displaystyle \frac{\Delta \widetilde{f}}{B}, \quad \widetilde{k}  =  \displaystyle \widetilde{m} \frac{2 \pi}{L} B $, and collecting  linear order terms of $\epsilon $, we obtain the non-dimensional equation in $\widetilde{f}$ as
\begin{align}
&{\rho_1}' {\rho_1}^2 {\rho}^2 \bigg[ {\rho_1'}^2 \alpha - 2 \log \left( \frac{\rho_1 \rho_1'}{\rho} \right) + 2 + \alpha \bigg] \widetilde{f}'' \nonumber \\
& + \rho_1 \rho \bigg[ {\rho_1'}^3 \rho_1 \alpha + [2 \rho_1 \rho_1'' \rho - \rho_1' \rho_1]~ 2 \log \left( \frac{\rho_1 \rho_1'}{\rho} \right) -  2 \rho_1'^2 \rho - 6 \rho_1 \rho_1'' \rho - 2 \rho_1 \rho_1'' \rho \alpha + 4 \rho_1' \rho_1 + \rho_1' \rho_1  \alpha  \bigg] \widetilde{f}'\nonumber \\
& -\rho_1' \bigg[ \rho_1'^2 \rho_1^2 \rho^2 \alpha \widetilde{k}^2 - 2 \log \left(\frac{\rho_1 \rho_1'}{\rho}\right) \rho_1'^2 \rho^2  + {\rho_1'}^2 \rho_1^2 \alpha + 4 {\rho_1'}^2 \rho^2 + {\rho_1'}^2 \rho^2 \alpha + 2 \rho_1 \rho_1'' \rho^2 - 2 \rho_1' \rho_1 \rho \bigg] \widetilde{f} = 0. \label{one_dim_axial_const_eq}
\end{align}
Eq. \eqref{one_dim_axial_const_eq} is subjected to internal pressure at inner boundary (at $\rho = A^*$) which is given as
\begin{align} 
&[2 \rho_1' A^*] \widetilde{f}  - \rho_1 \bigg[ -\rho_1'^2 A^* \alpha + 2 \log \left(\frac{\rho_1 \rho_1'}{A^*}\right) A^* + \rho_1 \wt{P} \rho_1' - 2 A^* - A^* \alpha \bigg] \widetilde{f}' = 0. \label{one_dim_inner_bound_const}
\end{align}
The boundary condition for the external constrained boundary (at $\rho = 1$ ) is 
\begin{align}
&\widetilde{f} = 0,\label{one_dim_const_bound}
\end{align}
and for unconstrained boundary is given by
\begin{align}
[2 \rho_1'] \widetilde{f} - \rho_1 \bigg[ -\rho_1'^2 \alpha + 2 \log \left({\rho_1 \rho_1'}\right) - 2 - \alpha \bigg] \widetilde{f}' = 0. \label{one_dim_free_bound}
\end{align}

\subsection{Solution using the compound matrix method}
\subsubsection{Condition for case 1}
The differential equations \eqref{non_dim_first_diff_eq} and \eqref{non_dim_sec_diff_eq} are converted into the system of first order linear differential equations in the form of $\mathbf{Y}'=\mathbf{AY}$ by substituting 
\begin{align}
f=y_1, ~  ~~ f'=y_2,~~~ g=y_3, ~~~ g'=y_4,
\end{align} 
which yields
\begin{align}
y_2' = -\frac{1}{a_1}\bigg[a_2 y_2 + a_3 y_1 +a_4 y_4 + a_5 y_3 \bigg], \quad \text{and} \quad y_4' = -\frac{1}{b_1} \bigg[b_2 y_4 + b_3 y_3 + b_4 y_2 + b_5 y_1 \bigg], \label{first_order_sys}
\end{align}
subjected to the boundary conditions \eqref{non_dim_bc_1} and \eqref{non_dim_bc_2}.
\noindent Now we convert the first order system $\mathbf{Y'}=\mathbf{H Y}$ into a new first order system of ODEs using compound variables in the form of $\mathbf{\Phi}'=\pmb{\mathcal{L}} \mathbf{\Phi}$ such that
\begin{equation}
\begin{aligned}
\Phi_1'&= {H}_{22} \Phi_1 + H_{23} \Phi_2 + H_{24} \Phi_3, \\
\Phi_2'&= \Phi_4 + \Phi_3, \\
\Phi_3'&= \Phi_5 + H_{42} \Phi_1 + H_{43} \Phi_2 + H_{44} \Phi_3,\\
\Phi_4'&=  \Phi_5 + H_{21} \Phi_2 + H_{22} \Phi_4 - H_{24} \Phi_{6}  ,\\
\Phi_5'&= H_{21} \Phi_3  + H_{22} \Phi_5 + H_{23} \Phi_6 - H_{41} \Phi_1 + H_{43} \Phi_4 + H_{44} \Phi_5,\\
\Phi_6'&=-H_{41} \Phi_2 - H_{42} \Phi_4 + H_{44} \Phi_6, 
\end{aligned} \label{compound_equations}
\end{equation}
where the coefficients in \eqref{compound_equations} are 
\begin{align*}
& H_{21} = - \frac{a_3}{a_1}, \quad H_{22} = - \frac{a_2}{a_1}, \quad H_{23} = - \frac{a_5}{a_1}, \quad H_{24} = - \frac{a_4}{a_1},\\
& H_{41} = - \frac{b_5}{b_1}, \quad H_{42} = - \frac{b_4}{b_1}, \quad H_{43} = - \frac{b_3}{b_1}, \quad H_{44} = - \frac{b_2}{b_1}.
\end{align*}

The boundary conditions \eqref{non_dim_bc_1} at $\rho=A^*$ are given by
\begin{equation} 
\begin{aligned}
a_{11} f + a_{22} f' + a_{33} g &=0, \\
b_{11} f + b_{44} g' &=0,
\end{aligned} \label{mod_bc1}
\end{equation} 
where the coefficients at the inner boundary ($\rho=A^*$) are defined as 
\begin{align*}
&a_{11} = 2 \rho_1' A^*, \quad a_{22}= - \rho_1 \bigg[ -\rho_1'^2 A^* \alpha +  A^* \left[ 2  \log \left( \frac{\rho_1 \rho_1'}{A^*}\right)  \right] + \rho_1 \rho_1' \wt{P} - 2 A^*  -  A^*  \alpha \bigg], \nonumber \\
 &  a_{33} = 2 \rho_1 \rho_1' A^* n, \quad  b_{11} = n \bigg [A^* \left[ 2 \log \left( \frac{\rho_1 \rho_1'}{A^*} \right) \right] +\rho_1 \rho_1' \wt{P} - A^* \alpha \bigg], \quad b_{44} = A^* \rho_1^2 \rho_1' \alpha .
\end{align*}

\subsubsection*{Initial condition}

The initial condition is given by using the compound variables
\begin{align}
\mbf{\Phi}(A^*)= \big[ \Phi_1, ~ \Phi_2,~ \Phi_3, ~ \Phi_4,~ \Phi_5,~ \Phi_6 \big] = \bigg[ -\frac{a_{33}}{a_{22}}, ~1, ~ 0, ~ -\frac{a_{11}}{a_{22}}, ~ - \frac{a_{33}}{a_{22}} \frac{b_{11}}{b_{44}},~ \frac{b_{11}}{b_{44}} \bigg] .\label{int_cond}
\end{align}
In order to obtain a non trivial solution the necessary condition for the objective function is $\det (\mbf{CM})=0$, where $\mbf{C}$ denotes the boundary condition \eqref{non_dim_bc_2} at $\rho = 1$ and $\mbf{M}$ denotes the solution matrix which are given as
\begin{align}
\mbf{C}=\begin{bmatrix}
1 & 0 & 0 & 0 \\
0 & 0 & 1 & 0 \\
\end{bmatrix},~~~~ \text{and}~~~~ \mbf{M}= \begin{bmatrix}
f_1 & f_2  \\ 
f_1'& f_2'  \\
g_1 & g_2\\
g_1' & g_2'
\end{bmatrix}. \label{sol_matrix}
\end{align}

\subsubsection{Conditions for case 2}
The boundary conditions of circumferentially perturbed cylinder associated with free outer surface is given by \eqref{non_dim_bc_free_case} and \eqref{non_dim_bc2_free_case} and corresponding target condition is obtained as
\begin{align}
\det(\mbf{CM}) = a_{11}^* b_{44}^* \Phi_{3} - a_{22}^* b_{11}^* \Phi_1 + a_{22}^* b_{44}^* \Phi_5 - a_{33}^* b_{11}^* \Phi_2 + a_{33}^* b_{44}^* \Phi_6 = 0 ,
\end{align}
where
\begin{align*}
a_{11}^* &= 2 \rho_1' , \quad  a_{22}^* = - \rho_1 \Bigg[ - \rho_1'^2 \alpha +  \left[ 2  \log \left( \frac{\rho_1 \rho_1'}{A^*}\right)  \right]  - 2   -   \alpha \Bigg], \quad  a_{33}^*  = 2 \rho_1 \rho_1'  n,\\
b_{11}^* &=  n \left[ 2 \log \left( \frac{\rho_1 \rho_1'}{1} \right) \right], \quad b_{44}^* = \rho_1^2 \rho_1' \alpha.
\end{align*}

\subsubsection{Conditions for case 3}
The initial condition (at $\rho = A^*$) of axially perturbed cylinder \eqref{diff_eq_non_dim_axial_const_bound} with constrained external surface is 
\begin{align}
\mbf{\Phi}(A^*) = \bigg[ -\frac{c_{33}^*}{c_{22}*}, ~1, ~ 0, ~ -\frac{c_{11}^*}{c_{22}^*}, ~ - \frac{c_{33}^*}{c_{22}^*} \frac{d_{11}^*}{d_{44}^*},~ \frac{d_{11}^*}{d_{44}^*} \bigg],
\end{align} 
and the target condition is $\det(\mbf{CM}) = 0$, where $\mbf{C}$ and $\mbf{M}$ corresponds to constrained boundary condition \eqref{axial_const_bc} at $\rho = 1$ and solution matrix, respectively. For this particular case, $\mbf{C}$ and $\mbf{M}$
are given as
\begin{align}
\mbf{C}=\begin{bmatrix}
1 & 0 & 0 & 0 \\
0 & 0 & 1 & 0 \\
\end{bmatrix},~~~~ \text{and}~~~~ \mbf{M}= \begin{bmatrix}
\bar{f}_1 & \bar{f}_2  \\ 
\bar{f}_1'& \bar{f}_2'  \\
\bar{h}_1 & \bar{h}_2\\
\bar{h}_1' & \bar{h}_2'
\end{bmatrix}. 
\end{align}

\subsubsection{Conditions for case 4}
The objective function or target condition for axially perturbed cylinder with unconstrained boundary condition is given as
\begin{align}
\det(\mbf{CM}) = c_{111}^* d_{444}^* \Phi_{3} - c_{222}^* d_{111}^* \Phi_1 + c_{222}^* d_{444}^* \Phi_5 - c_{333}^* d_{111}^* \Phi_2 + c_{333}^* d_{444}^* \Phi_6 = 0,
\end{align}
where
\begin{align*}
c_{111}^* &= 2 \rho_1' , \quad  c_{222}^* = -\rho_1 \bigg[ -\rho_1'^2 \alpha + 2 \log \left(\rho_1 \rho_1' \right)  - 2 - \alpha \bigg], \quad  c_{333}^*  = 2 \rho_1 \rho_1' k,\\
d_{111}^* &=   k \bigg[ 2 \log \left(\rho_1 \rho_1' \right) - \alpha \bigg], \quad d_{444}^* =    \alpha \rho_1' .
\end{align*}

\section{Appendix: Description of the numerical technique} \label{numerical_tech_app}
We describe the compound matrix method \citep{haughton1997eversion, haughton2008evaluation, Mehta2021} and shooting method \citep{haughton1979bifurcation2, saxena2018finite} for the solution of ODEs
\subsection*{Compound matrix method}
Equations \eqref{non_dim_first_diff_eq} and \eqref{non_dim_sec_diff_eq} can be written as two-point boundary value problem expressed in first order ODEs
\begin{align}
\frac{d \mathbf{Y}}{dX}= {\mbf{H}(\lambda,x)} \mathbf{Y}, \hspace{0.8in} x \in (a,b), \label{BVP_diff_eqn}
\end{align} 
subjected to boundary conditions
\begin{align}
\mathbf{BY}&=\mathbf{0}, \hspace{1in} x=a, \nonumber\\
\mathbf{CY}&=\mathbf{0}, \hspace{1in} x=b, 
\end{align}
where $\lambda$ is the eigenvalue or critical buckling parameter, $\mathbf{Y}$ is $1 \times 2q$ vector, ${\mbf{H}}$ is $2q \times 2q$ matrix and $\mathbf{B}$ and $\mathbf{C}$ both are $q \times 2q $ full rank matrices i.e., $q$ boundary conditions are given at  
$x=a,b$. Assume the general solution of \eqref{BVP_diff_eqn} is in the form of
\begin{align}
    \mathbf{y}(\lambda,x)= \sum_{j=1}^{q} p_j \mathbf{y}^{j},
\end{align}
where $\mbf{y} = \{\mathbf{y}^{(1)} (\lambda,x), \mathbf{y}^{(2)} (\lambda, x), ..., \mathbf{y}^{(q)}(\lambda, x)\} $ is a set of $q$ linear independent solution of \eqref{BVP_diff_eqn} and $p_1, p_2, ..., p_q$ are the constants. Solution matrix $\mathbf{M}$ to be $2q \times q$ is define as $\mathbf{M} = [\mathbf{y}^{(1)}, \mathbf{y}^{(2)},  ... \mathbf{y}^{(q)} ]$, and \eqref{BVP_diff_eqn} in terms of $\mathbf{M}$ is given by
\begin{align}
\frac{d \mathbf M}{dx}=[{\mbf{A}} \mathbf{y}^{(1)}, {\mbf{A}} \mathbf{y}^{(2)}, . . .,{\mbf{A}} \mathbf{y}^{(q)} ]= {\mbf{A}} \mathbf{M}. 
\end{align}
The compound variables are defined as minors of $\mathbf{M}$ denoted as $\Phi_1, \Phi_2, ...,$ and those are  ${2q \choose q}$ in numbers. In this current work, Eqs. \eqref{first_order_sys} and \eqref{BVP_diff_eqn} is a fourth order ODE system ($q=2$) for which the solution matrix is
\begin{align}
\mathbf{M}=\begin{bmatrix}
y_1^{(1)} & y_1^{(2)}  \\ 
y_2^{(1)} & y_2^{(2)} \\
y_3^{(1)} & y_3^{(2)} \\
y_4^{(1)} & y_4^{(2)} 
\end{bmatrix} = \begin{bmatrix}
f_1 & f_2  \\ 
f_1'& f_2'  \\
g_1 & g_2\\
g_1' & g_2'
\end{bmatrix},
\end{align}
and 6 minors of $\mathbf{M}$

\begin{align}
\Phi_1& = (1,2)=  \begin{vmatrix}
y_1^{(1)} & y_1^{(2)} \\
y_2^{(1)} & y_2^{(2)}
\end{vmatrix}, \hspace{0.5 in} \Phi_2=(1,3)=  \begin{vmatrix}
y_1^{(1)} & y_1^{(2)} \\
y_3^{(1)} & y_3^{(2)}
\end{vmatrix}, \nonumber \\[3pt]
\Phi_{3}&=(1,4), ~~ \Phi_{4}=(2,3),  ~~ \Phi_{5}=(2,4), ~~ \Phi_{6}=(3,4). 
\end{align}
The system is now converted into ${2q \choose q}$ ODEs which is in the form of
\begin{align}
\mathbf{\Phi'}= \pmb{\mathcal{L}} \mathbf{\Phi}, \hspace{1in} \rho \in (A^*,B),  \label{six_order_cmp_system}
\end{align} 
where
\begin{align*}
\Phi_1'=\begin{vmatrix}
y_1^{(1)} & y_1^{(2)}  \\ 
y_2^{(1)} & y_2^{(2)}
\end{vmatrix}'& = \begin{vmatrix}
{y_1^{(1)}}' & {y_1^{(2)}}'  \\ 
{y_2^{(1)}} & {y_2^{(2)}} 
\end{vmatrix} +
\begin{vmatrix}
{y_1^{(1)}} & {y_1^{(2)}}  \\ 
{y_2^{(1)}}' & {y_2^{(2)}}' 
\end{vmatrix}, \nonumber \\
&=\begin{vmatrix}
\sum_{j=1}^{4} {H}_{1j} y_j^{(1)} & \sum_{j=1}^{4} {H}_{1j} y_j^{(2)} \\
{y_2^{(1)}} & {y_2^{(2)}}
\end{vmatrix} + \begin{vmatrix}
y_1^{(1)} & y_1^{(2)}  \nonumber\\
\sum_{j=1}^{4} {H}_{2j} y_j^{(1)} & \sum_{j=1}^{4} {H}_{2j} y_j^{(2)}
\end{vmatrix}, \nonumber\\
& = {H}_{22} \Phi_1 + {H}_{23} \Phi_{2} +{H}_{24} \Phi_3,
\end{align*} 
and the remaining equations are given in \eqref{compound_equations}.
The initial condition associated with the system of six ODEs \eqref{six_order_cmp_system} at $\rho = A^*$ is 
\begin{align} 
\mathbf{\Phi}(A^*) = [\Phi_1,~ \Phi_2,~ \Phi_3,~ \Phi_4,~ \Phi_5,~ \Phi_6]. \label{initial_cond_phi}
\end{align} 
Initial condition is evaluated using boundary condition \eqref{mod_bc1} which is rewritten as
\begin{equation}
\begin{aligned}
f' = - \frac{1}{a_{22}} \bigg[ a_{11} f + a_{33} g \bigg], \qquad \text{and} \qquad g' = -\frac{b_{11}}{b_{44}}f.
\end{aligned} \label{rewrite_BC_const}
\end{equation}

\noindent Using \eqref{rewrite_BC_const} and \eqref{sol_matrix}, the matrix entries in \eqref{initial_cond_phi} are evaluated as   
\begin{align*}
\Phi_1 &= \begin{vmatrix}
f_1 & f_2 \\
f_1' & f_2'
\end{vmatrix} = \begin{vmatrix}
f_1 & f_2 \\
\displaystyle - \frac{1}{a_{22}} \big[ a_{11} f_1 + a_{33} g_1 \big] & \displaystyle - \frac{1}{a_{22}} \big[ a_{11} f_2 + a_{33} g_2 \big] 
\end{vmatrix} = -\frac{a_{33}}{a_{22}} \begin{vmatrix}
f_1 & f_2 \\
g_1 & g_2
\end{vmatrix}, \\
\Phi_2 &= \begin{vmatrix}
f_1 & f_2 \\
g_1 & g_2
\end{vmatrix}, \qquad 
\Phi_3 = \begin{vmatrix}
f_1 & f_2 \\
g_1' & g_2'
\end{vmatrix}  = \begin{vmatrix}
f_1 & f_2 \\
-\displaystyle \frac{b_{11}}{b_{44}} f_1 &  -\displaystyle \frac{b_{11}}{b_{44}} f_2
\end{vmatrix} = 0, \qquad 
\Phi_4 = \begin{vmatrix}
f_1' & f_2' \\
g_1 & g_2
\end{vmatrix} = 
-\frac{a_{11}}{a_{22}} \begin{vmatrix}
f_1 & f_2 \\
g_1 & g_2
\end{vmatrix}, \\
\Phi_5 &= \begin{vmatrix}
f_1' & f_2' \\
g_1' & g_2'
\end{vmatrix} = -\displaystyle \frac{a_{33}}{a_{22}} ~ \frac{b_{11}}{b_{44}} 
\begin{vmatrix} 
f_1 & f_2 \\
g_1 & g_2
\end{vmatrix}, \qquad 
\Phi_6 = \begin{vmatrix}
g_1 & g_2 \\
g_1' & g_2'
\end{vmatrix} =  \frac{b_{11}}{b_{44}} \begin{vmatrix}
f_1 & f_2 \\
g_1 & g_2
\end{vmatrix}.
\end{align*}
Now, if we assume $\Phi_2 = 1 $, then 
\begin{align}
\Phi_1 = -\frac{a_{33}}{a_{22}}, \quad \Phi_{2} = 1, \quad \Phi_{3} = 0, \quad \Phi_{4} = -\frac{a_{11}}{a_{22}}, \quad \Phi_5 = - \frac{a_{33}}{a_{22}} \frac{b_{11}}{b_{44}}, \quad \Phi_6 = \frac{b_{11}}{b_{44}}.
\end{align} 
The initial condition is
\begin{align}
\mbf{\Phi}(A^*) = \bigg[ -\frac{a_{33}}{a_{22}}, ~1, ~ 0, ~ -\frac{a_{11}}{a_{22}}, ~ - \frac{a_{33}}{a_{22}} \frac{b_{11}}{b_{44}},~ \frac{b_{11}}{b_{44}} \bigg]. \label{final_int_cond}
\end{align}
The system of equations \eqref{six_order_cmp_system} is now numerically integrated using initial condition \eqref{final_int_cond} which produces the solution $y^{(j)}$ at $\rho = 1$
\begin{align}
\mathbf{Cy}=\mathbf{C} \sum_{j=1}^{q} p_j \mathbf{y}^{(j)}(b)=\mathbf{CMp}=\mathbf{0}.
\end{align}
Necessary condition for existence of non-trivial solution of Eq. \eqref{six_order_cmp_system} is
\begin{align}
\det(\mathbf{CM})=0.\label{A_13}
\end{align} 

\subsection*{Description of Shooting method for constrained boundary}
The linear system of equation \eqref{first_order_sys} is rewritten as
\begin{equation}
\begin{aligned}
a_3y_1 + a_2 y_2 + a_4 y_4 + a_5 y_3 + a_1 y_2' &= 0, \\
b_5 y_1 + b_4 y_2 + b_3 y_3 + b_2 y_4 + b_1 y_4' &=0. \label{shoot_diff_eq}
\end{aligned}
\end{equation}
where $ [y_1, \ y_2, \ y_3, \ y_4] = [f,\  f',\  g,\  g'] $. 
The system of first order ODEs using \eqref{shoot_diff_eq} is given by
\begin{align}
\mbf{Dy}' = \mbf{g},  \label{sys_in_simple_shoot}
\end{align} 
where 
\begin{align*}
\mbf{D} =  \begin{bmatrix}
1 & 0 & 0 & 0\\
0 & a_1 & 0 & 0\\
0 & 0 & 1 & 0 \\
0 & 0 & 0 & b_1
\end{bmatrix}, \quad \mbf{y}' = \begin{bmatrix}
y_1' \\
y_2'\\
y_3'\\
y_4'
\end{bmatrix}, \quad \mbf{g} = \begin{bmatrix}
y_2 \\
-a_3 y_1 - a_2 y_2 - a_5 y_3 - a_4 y_4\\
y_4\\
-b_5 y_1 - b_4 y_2 - b_3 y_3 -b_2 y_4
\end{bmatrix}.
\end{align*}
We convert this system \eqref{sys_in_simple_shoot} into initial value problem with general initial conditions
\begin{align}
y_i^{(j)} \bigg|_{r=A} = \delta_{ij}. \label{gen_sol_shoot}
\end{align}
Here, $i = 1,2,..., 4$ for each set $j = 1,2,..., 4$ which makes the initial condition for each set to be
\begin{align}
\mbf{y}^{(1)} = [1 ~ 0 ~ 0 ~ 0], \quad \mbf{y}^{(2)} = [0 ~ 1 ~ 0 ~ 0], \quad \mbf{y}^{(3)} = [0 ~ 0 ~ 1 ~ 0], \quad \mbf{y}^{(4)} = [0 ~ 0 ~ 0 ~ 1].
\end{align}
The general solution is assumed to be the linear combination of obtained solution such as
\begin{align}
{y}_i = \sum_{j=1}^{4} c_j~ {y}_i^{(j)}.
\end{align} 
The boundary conditions are given by \eqref{mod_bc1}.
Upon substituting the general solution \eqref{gen_sol_shoot} in \eqref{mod_bc1}, we obtain the system of linear algebraic equations as
\begin{subequations} 
\begin{align}
&a_{11} \big[ c_1 y_1^{(1)} + c_2 y_1^{(2)} + c_3 y_1^{(3)} + c_4 y_1^{(4)}\big] + a_{22}\big[ c_1 y_2^{(1)} + c_2 y_2^{(2)} + c_3 y_2^{(3)} + c_4 y_2^{(4)}\big] \\ \nonumber
& + a_{33} \big[ c_1 y_3^{(1)} + c_2 y_3^{(2)} + c_3 y_3^{(3)} + c_4 y_3^{(4)}\big] = 0, \\
& b_{11} \big[ c_1 y_1^{(1)} + c_2 y_1^{(2)} + c_3 y_1^{(3)} + c_4 y_1^{(4)}\big] + b_{44} \big[ c_1 y_4^{(1)} + c_2 y_4^{(2)} + c_3 y_4^{(3)} + c_4 y_4^{(4)}\big] =0,\\
& c_1 y_1^{(1)} + c_2 y_1^{(2)} + c_3 y_1^{(3)} + c_4 y_1^{(4)} = 0,\\
& c_1 y_3^{(1)} + c_2 y_3^{(2)} + c_3 y_3^{(3)} + c_4 y_3^{(4)}=0.
\end{align} \label{non_trivial_sys}
\end{subequations}
The system \eqref{non_trivial_sys} is in the form of $Z_{ek} c_k =0$ where the matrix $\mbf{Z}$ is given by 
\begin{align*}
\fontsize{10}{10}{
\mbf{Z} = \begin{bmatrix}
a_{11} y_1^{(1)} + a_{22} y_2^{(1)} + a_{33} y_3^{(1)} & a_{11} y_1^{(2)} + a_{22} y_2^{(2)} + a_{33} y_3^{(2)}  &  a_{11} y_1^{(3)} + a_{22} y_2^{(3)} + a_{33} y_3^{(3)} & a_{11} y_1^{(4)} + a_{22} y_2^{(4)} + a_{33} y_3^{(4)} \\
b_{11} y_1^{(1)} + b_{44} y_4^{(1)} & b_{11} y_1^{(2)} + b_{44} y_4^{(2)} & b_{11} y_1^{(3)} + b_{44} y_4^{(3)} & b_{11} y_1^{(4)} + b_{44} y_4^{(4)}\\
y_1^{(1)} & y_1^{(2)} & y_1^{(3)} & y_1^{(4)} \\
y_3^{(1)} & y_3^{(2)} & y_3^{(3)} & y_3^{(4)}
\end{bmatrix}.
}
\end{align*}
For non-trivial solution of such system $\det(\mbf{Z})$ vanishes.

\section{Appendix: Fibres in axial direction} \label{Fibers_app}
The soft hyperelastic cylinder is made anisotropic by a reinforcement of fibres orientated along the vector $\mbf{a}$. To account for this reinforcement, the strain energy density function \eqref{strain_energy} has additional fibre terms as
\begin{align}
 \Omega^*(I_1, I_3,I_4) = \frac{\mu}{2} \big[ I_1 - 3 - \log I_3 \big] + \frac{\kappa}{4} \big[ \log I_3 \big]^2 + \Omega_{f}(I_4),
\end{align}
where $\Omega_f(I_4)= \displaystyle \frac{k_1}{2 k_2} \bigg[ \exp[k_2 [I_4 - 1]^2] - 1 \bigg]$ is the energy due to fibre reinforcement. The first Piola Kirchhoff stress corresponding to fibres reinforcement in axial direction is 
\begin{equation}
\begin{aligned}
\mbf{P}_f = \frac{\partial \Omega_f}{\partial \mbf{F}} &= \frac{\partial \Omega_f}{\partial I_4} \frac{ \partial I_4}{\partial \mbf{F}}= \Bigg[k_1 [I_4-1]  \exp \big[k_2 [I_4 - 1]^2 \big]\Bigg] \bigg[2 \mbf{a} \otimes \mbf{Fa}\bigg],
\end{aligned}
\end{equation}
where $\mbf{a}$ denote the unit vector which characterized the direction of fibers. The elastic moduli corresponding to fibre term is given by 
\begin{align}
\pmb{\mathcal{A}}^f  =  \frac{\partial \mbf{P}_f}{\partial \mbf{F}} = & k_1 \bigg[\frac{\partial I_4}{ \partial \mbf{F}}\bigg] \exp \big[k_2 [I_4 - 1]^2 \big] \bigg[2 \mbf{a} \otimes \mbf{Fa}\bigg] + k_1 [I_4 -1] \frac{\partial }{\partial \mbf{F}} \bigg[ \exp \big[k_2 [I_4 - 1]^2\big] \bigg] \bigg[2 \mbf{a} \otimes \mbf{Fa}\bigg] \nonumber \\
&  + k_1 [I_4 -1] \exp \big[k_2 [I_4 - 1]^2 \big] \frac{\partial }{\partial \mbf{F}} \left( 2 \mbf{a} \otimes \mbf{Fa} \right),
\end{align}
which is written in index notation as
\begin{align}
\mathcal{A}^{f}_{ijkl} = & 2 \Bigg[k_1  \exp \big[k_2 [I_4 - 1]^2 \big] \bigg[ \big[\mbf{a} \otimes \mbf{Fa} \big]_{ij} \big[\mbf{a} \otimes \mbf{Fa} \big]_{kl} \bigg] \nonumber\\
& + 2 k_1 [I_4 - 1] \exp \big[k_2 [I_4 - 1]^2 \big] 2 k_2 [I_4 - 1] \bigg[ \big[\mbf{a} \otimes \mbf{Fa} \big]_{ij} \big[\mbf{a} \otimes \mbf{Fa} \big]_{kl} \bigg] \nonumber\\
& + 2 k_1 [I_4 - 1] \exp \big[k_2 [I_4 - 1]^2 \big] a_i a_j \delta_{kl} \Bigg]. \label{index_moduli_fibre}
\end{align}
Upon simplifying \eqref{index_moduli_fibre} we obtain
\begin{align}
\mathcal{A}^{f}_{ijkl} =  2 k_1 \exp \big[k_2 [I_4 - 1]^2 \big]  \Bigg[ \bigg[ 1 + 2 k_2 [I_4 - 1] \bigg] \big[\mbf{a} \otimes \mbf{Fa} \big]_{ij} \big[\mbf{a} \otimes \mbf{Fa} \big]_{kl} + [I_4 - 1] a_i a_j \delta_{kl} \Bigg]. \label{simp_index_moduli}
\end{align}
The incremental stress associated with the fibre term is 
\begin{align}
\delta \mbf{P}_f = \pmb{\mathcal{A}}^f \delta \mbf{F} = \mathcal{A}^{f}_{ijkl} [\delta {\mbf{F}}]_{kl}. \label{fib_inc_stress}
\end{align} 
This can be expanded by substituting \eqref{simp_index_moduli} in \eqref{fib_inc_stress} to get
\begin{align}
\delta \mbf{P}_f =  & 2 k_1 \exp \big[k_2 [I_4 - 1]^2 \big] \bigg[ 1 + 2 k_2 [I_4 - 1] \bigg] \big[\mbf{a} \otimes \mbf{Fa} \big] \bigg[ \text{tr} \bigg( \delta \mbf{F}^T [\mbf{a} \otimes \mbf{Fa}] \bigg)\bigg] \nonumber \\
& + 2 k_1 [I_4 - 1] \exp \big[k_2 [I_4 - 1]^2 \big] \left[\mbf{a} \otimes \mbf{a} \right] \text{tr}(\delta \mbf{F}^T).
\end{align}
\end{appendix}
\end{document}